\documentclass[sigplan,screen, 10pt]{acmart}

\usepackage{tikz}
\usepackage{amsmath}
\usepackage[utf8]{inputenc}
\usepackage{graphicx}
\usepackage{caption}
\usepackage{subcaption}
\usepackage{comment}
\usepackage{xspace}
\usepackage{enumitem}
\usepackage{multirow}

\newcommand{\system}{\texttt{RPS-Serve}\xspace}

\newcommand{\takeaways}{\noindent\textit{\textbf{Takeaways: }}}

\usepackage{pifont}

\newenvironment{tightitemize}%
	 {\begin{list}{$\bullet$}{%
 		\setlength{\leftmargin}{10pt}
        \setlength{\itemsep}{0pt}%
        \setlength{\parsep}{0pt}%
        \setlength{\topsep}{0pt}%
        \setlength{\parskip}{0pt}%
        }%
    }%
    {\end{list}}

  {\begin{list}{\arabic{enumi}.}{%
        \usecounter{enumi}%
        \setlength{\leftmargin}{15pt}%
        \setlength{\itemsep}{0pt}%
        \setlength{\parsep}{0pt}%
        \setlength{\topsep}{0pt}%
        \setlength{\parskip}{0pt}%
    }%
  }%
  {\end{list}}

\newcommand{\custompar}[1]{\noindent{\textbf{#1.}}}
\newcommand{\customsection}[1]{\noindent{\textit{\textbf{#1.}}}}

\copyrightyear{2026}\acmYear{2026}
\setcopyright{cc}
\setcctype{by}
\acmConference[SIGOPS ATC 2026]{The 2026 ACM SIGOPS Annual Technical Conference}{November 15-18, 2026}{Hong Kong, China}
\acmBooktitle{The 2026 ACM SIGOPS Annual Technical Conference (SIGOPS ATC 2026), November 15-18, 2026, Hong Kong, China}
\acmDOI{10.1145/3828523.3857088}
\acmISBN{979-8-4007-2818-1/2026/11}

\begin{document}
%
%


\graphicspath{ {./figures/} }

\title[Rocks, Pebbles and Sand: Modality-aware Scheduling for MLLM Inference]{Rocks, Pebbles and Sand: Modality-aware Scheduling for Multimodal Large Language Model Inference}

\author{Konstantinos Papaioannou}
\orcid{0009-0003-0573-3254}
\affiliation{%
  \institution{IMDEA Software Institute}
  \institution{Universidad Politécnica de Madrid}
  \city{}
  \country{}}
\email{konstantinos.papaioannou@imdea.org}

\author{Thaleia Dimitra Doudali}
\orcid{0000-0002-3197-839X}
\affiliation{%
  \institution{IMDEA Software Institute}
  \city{Madrid}
  \country{Spain}}
\email{thaleia.doudali@imdea.org}

\begin{abstract}

Multimodal Large Language Models (MLLMs) power platforms like ChatGPT, Gemini, and Copilot, enabling rich interactions across text, images, and videos.
Unlike text-only workloads, multimodal requests are inherently heterogeneous, differing by orders of magnitude in latency and memory footprint.
Existing LLM serving systems, designed for homogeneous text workloads, struggle under this heterogeneity: large image and video requests monopolize GPU resources, causing severe head-of-line blocking and degrading the responsiveness of lightweight, latency-critical text queries.
Recent multimodal serving systems mitigate this interference through disaggregation, isolating encoding and inference stages across GPUs.
While effective at large scale, such designs introduce severe GPU underutilization in more practical, moderate-scale deployments. 

In response, we build RPS-Serve a modality-aware scheduling framework that leverages the extreme heterogeneity in multimodal workloads and introduces a simple abstraction: videos behave like rocks, images like pebbles, and text like sand.
Then, it lets sand flow quickly through pebbles and rocks, maximizing GPU utilization while ensuring fast responsiveness for lightweight requests and controlled execution for heavier ones.
To achieve this, RPS-Serve estimates the GPU admission cost of each request, classifies requests into rocks, pebbles, and sand using lightweight clustering, and applies priority-based scheduling with aging to balance performance across modalities.
Evaluation across state-of-the-art MLLMs shows that RPS-Serve increases GPU utilization by 65\% and reduces TTFT by up to 9.3$\times$, on average across models, compared to disaggregation-based multimodal systems.

\end{abstract}



\settopmatter{printfolios=false,printacmref=true} 
\maketitle

\section{Introduction}
\label{sec:intro}

Large Language Models (LLMs) have revolutionized natural language processing by enabling tasks such as text generation, summarization, and reasoning at scale, becoming the backbone of applications from conversational agents to code assistants.
However, as user interactions increasingly involve richer content, such as images, videos, and audio, the paradigm is shifting.
Today, widely used platforms such as ChatGPT~\cite{chatgpt} and integrated models like Gemini~\cite{gemini-google} and Copilot~\cite{copilot} already employ \textbf{Multimodal Large Language Models (MLLMs)} to serve users.
MLLMs process diverse modalities alongside text, unlocking capabilities such as image reasoning, video summarization, and audio captioning, while preserving the interactive nature of traditional LLMs.
Examples of MLLMs include GPT-4~\cite{gpt4}, Chameleon~\cite{chameleon}, and open-source models like LLaVA~\cite{liu2023llava}.
In addition, emerging ``any-to-any'' models such as Next-GPT~\cite{wu24next} go beyond text outputs, enabling responses in multiple modalities (e.g., generating images or videos), further expanding the scope of multimodal AI.

\begin{figure}[t]
    \centering
    \includegraphics[width=0.92\linewidth]{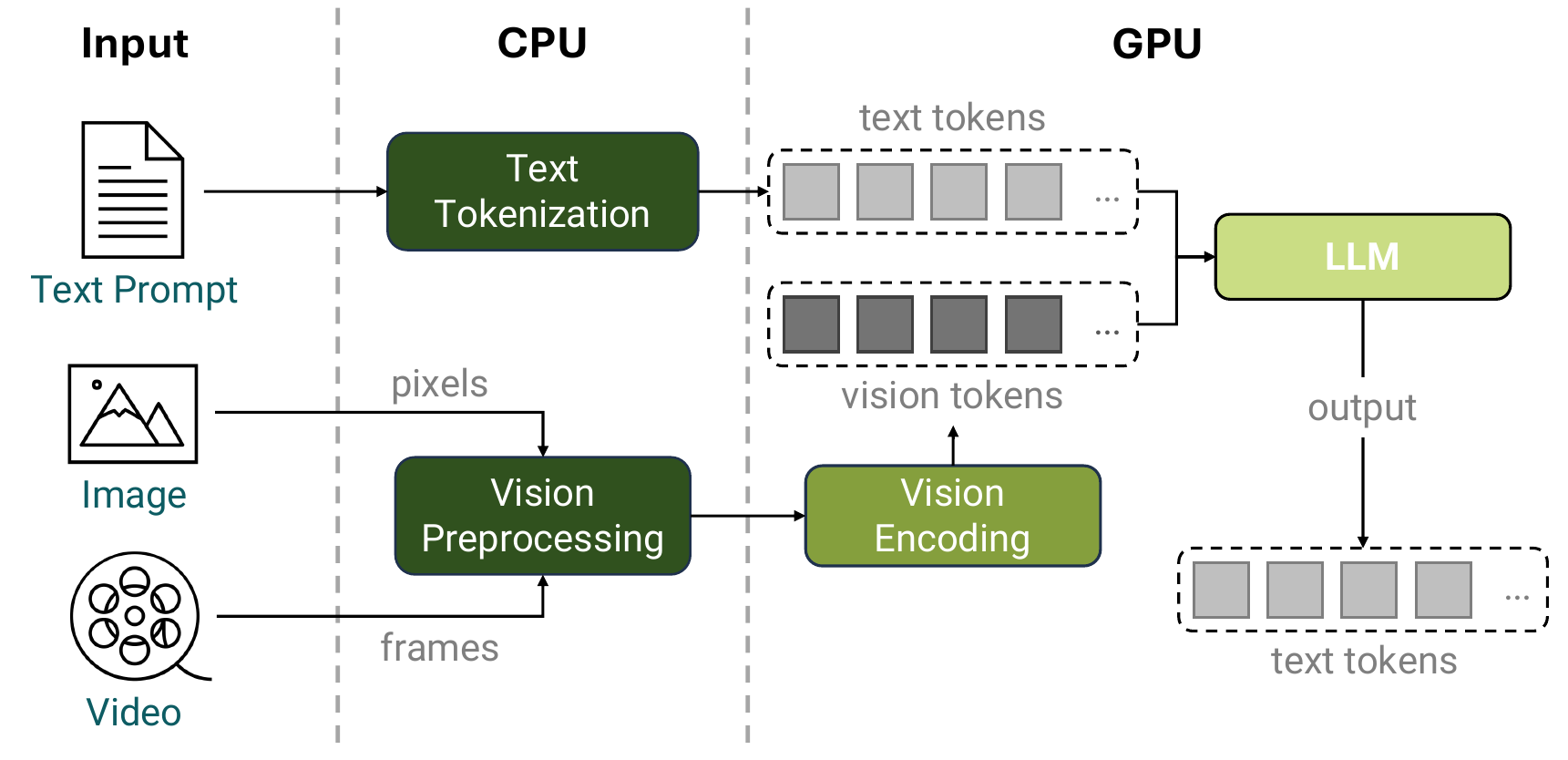}
    \caption{Multimodal LLM (MLLM) Inference Stages.}
    \label{fig:mllm-arch}
    \vspace{-0.1in}
\end{figure}

Unlike traditional LLMs, which process only text, MLLMs introduce additional inference stages for non-text modalities.
As shown in Figure~\ref{fig:mllm-arch}, multimodal inputs undergo vision \emph{preprocessing} and \emph{encoding} before reaching the LLM backend, increasing both latency and memory footprint.
Our characterization (Section~\ref{ssec:characterization-iso}) reveals substantial differences across modalities.
Text requests, while highly variable in length~\cite{sharegpt,alpaca,lmsyschat1m,servegen,longbench}, remain lightweight.
In contrast, image and video requests consume {\it orders of magnitude more memory}, making them significantly more resource-intensive; videos further amplify this effect as bursts of image frames processed together.
As a result, video requests dominate GPU usage, followed by images, while text remains minimal.
Latency follows a similar behavior.
Visual modalities also exhibit heavy-tailed behavior, where a subset of requests incurs significantly higher latency.
These observations align with recent production-scale multimodal workload characterizations~\cite{servegen, modserve}.
Overall, multimodal workloads are heterogeneous, bursty, and heavy-tailed, making them {\it fundamentally distinct from homogeneous text workloads}.

Despite these clear differences, current state-of-the-art LLM serving systems, such as vLLM~\cite{vllm,paged-attention}, are not designed for multimodal heterogeneity.
Their default first-come-first-served (FCFS) scheduling policy, while simple and efficient for homogeneous text workloads, performs poorly under multimodal settings. 
Large image and video requests incur expensive prefills due to both modality-specific encoding and longer LLM prompt processing, monopolizing GPU resources and causing severe head-of-line (HoL) blocking for subsequent requests.
As a result, lightweight text requests can experience severe delays, violating their tight latency targets~\cite{gpt4-sla, dist-serve, dynamollm}.
Similar challenges have been observed in text-only LLM serving, where prompts with long input contexts delay shorter ones during prefill~\cite{longbench,sarathi-serve,sarathi,loongserve,mnemosyne}.
To mitigate such interference, prior LLM serving systems~\cite{dist-serve, splitwise, mooncake} introduce \emph{disaggregation}, separating prefill and decode execution across dedicated GPUs to reduce HoL delays.

Since multimodal workloads further exacerbate HoL delays, recent multimodal serving systems~\cite{modserve, elasticmm, epd} also leverage disaggregation to mitigate delays from the additional vision encoding stage.
These systems vary in how they decompose the pipeline: some isolate only the vision encoder onto a separate GPU~\cite{modserve}, others fully disaggregate execution into encoding, prefill, and decode stages~\cite{epd}, while more general designs also separate text-only and multimodal workloads into distinct execution paths~\cite{elasticmm}.
While effective at large scale (e.g., deployments with up to 128 GPUs~\cite{modserve}), where abundant resources allow dedicated allocation per inference stage, these approaches face limitations in more practical settings.
In moderate-scale deployments (e.g., 4-GPU servers), they introduce a fundamental {\it trade-off between performance and resource efficiency}.
By dedicating different GPUs across stages, disaggregation leads to substantial GPU underutilization (Section~\ref{ssec:limitations-of-sota}) and diminishes the overall performance benefits.

Therefore, while disaggregation reduces head-of-line (HoL) blocking in multimodal workloads, it suffers from resource underutilization.
At the same time, existing LLM serving systems (e.g., vLLM) improve utilization but remain vulnerable to HoL blocking due to FCFS scheduling.
This exposes a fundamental gap: current serving systems either prioritize performance or resource efficiency, but not both.
Instead, in this paper we ask: \emph{can a scheduling approach for multimodal inference simultaneously mitigate head-of-line blocking and maintain high resource utilization?}

To answer this question, we present {\bf \system}, a modality-aware scheduling framework for multimodal LLM inference.
\system is based on the insight that multimodal requests differ by orders of magnitude in both time and memory, mapping them to a simple abstraction: video requests behave like \emph{rocks}, dominating GPU resources; image requests are \emph{pebbles}, moderately heavy; and text-only requests are \emph{sand}, lightweight yet latency-sensitive.
\system lets {\it sand flow quickly through pebbles and rocks}, maximizing GPU utilization while balancing fast responsiveness for sand with controlled execution of heavier pebble and rock requests. 

To do so, \system estimates the GPU admission cost of each request using its prefill latency and memory footprint, and classifies requests into sand, pebbles, and rocks using lightweight clustering.
It maintains separate queues per class and performs priority-based scheduling enriched with an aging mechanism.
At every scheduling iteration, \system dynamically selects the requests that will execute in the batch based on these priorities, adapting to workload heterogeneity while balancing responsiveness and resource efficiency.
Evaluation across state-of-the-art MLLMs shows that \system increases GPU utilization by 65\% and reduces TTFT by up to 9.3$\times$, on average across models, compared to disaggregation-based multimodal systems.

\noindent The specific {\bf paper contributions} are:
\begin{tightitemize}
    \item A detailed characterization of multimodal LLM inference workloads and analysis of the limitations of current multimodal serving systems (Section~\ref{sec:motivation}).
    \item The design of \system \footnote{\system is available at: \url{https://github.com/muse-research-lab/rps-serve}, \url{https://zenodo.org/records/23134685}}, an {\bf open-source} modality-aware scheduling framework for multimodal inference (Section~\ref{sec:system}).
    \item A comprehensive evaluation of \system on state-of-the-art multimodal serving systems and models (Section~\ref{sec:evaluation}).
\end{tightitemize}
\section{Motivation}
\label{sec:motivation}

First, this section provides background on the architecture and inference stages of multimodal LLMs (Section~\ref{ssec:mllm-arch}).
Then, it characterizes the performance and memory behavior of multimodal requests across different models (Section~\ref{ssec:characterization-iso}).
Finally, it quantifies the limitations of current state-of-the-art multimodal serving systems (Section~\ref{ssec:limitations-of-sota}).

\subsection{Multimodal LLM Architecture}
\label{ssec:mllm-arch}

Figure~\ref{fig:mllm-arch} shows the internal components of a decoder-only multimodal LLM.
During inference, the model processes the input data modality (e.g., image, video, etc.) along with the accompanying text question. 
For example, a user may upload an image of a city street along with the text prompt “Describe the architectural style of the buildings in this photo”. 
The multimodal data is first preprocessed into intermediate representations (e.g., pixels for images, frames for videos, etc.) and then encoded into embeddings, that capture the semantics of these modalities in a unified format.
The text is first split in smaller parts (tokens) and then each token is mapped into an embedding according to the predefined vocabulary.

Internally, the multimodal LLM uses an unmodified LLM backend that processes these embeddings into two distinct inference phases: the {\it prefill} (or prompt) phase and the {\it decode} phase, which spans multiple iterations~\cite{orca}.
In the prefill phase, the input prompt is processed all together, and during the decode phase the LLM generates the output text one token at a time in an auto-regressive manner, with each new token relying on the previously generated ones.
While this phase is not as compute intensive as prefill, it is memory-intensive due to the use of the {\it KV Cache}, which caches the initial prompt and the generated tokens so far, to avoid recomputation.
The KV Cache can grow significantly in size and reach the GPU memory limits~\cite{attentionstore, ragcache, cachegen}.

\subsection{Characterization in Isolation}
\label{ssec:characterization-iso}

\begin{figure}[t]
    \centering
    \includegraphics[width=0.5\linewidth]{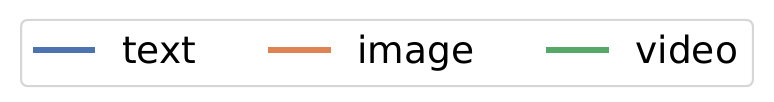}
    
    \begin{subfigure}[t]{\columnwidth}
        \centering
        \includegraphics[width=\linewidth]{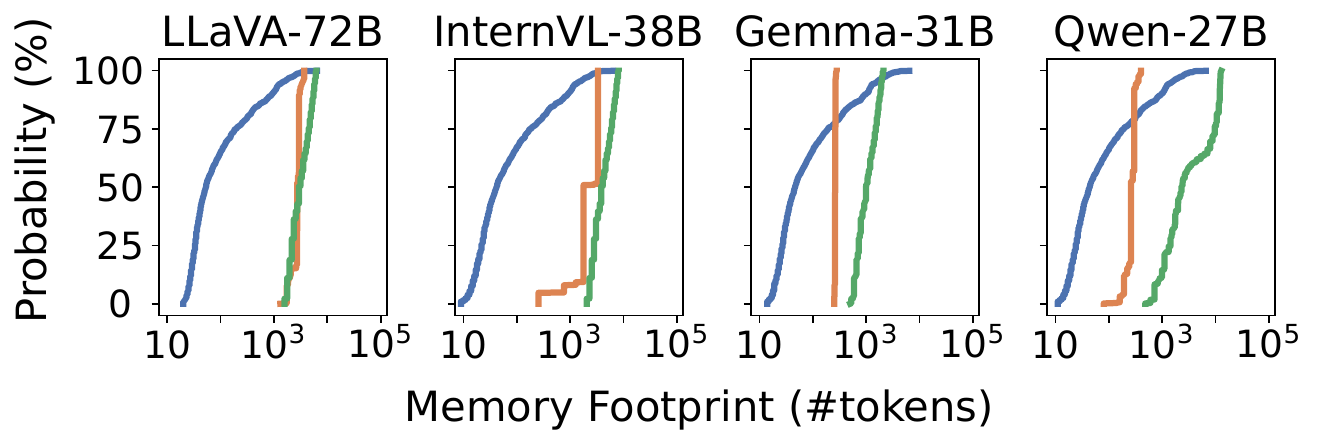}
        \caption{Memory Footprint}
        \label{fig:mem-cdf}
    \end{subfigure}
    
    \begin{subfigure}[t]{\columnwidth}
        \centering
        \includegraphics[width=\linewidth]{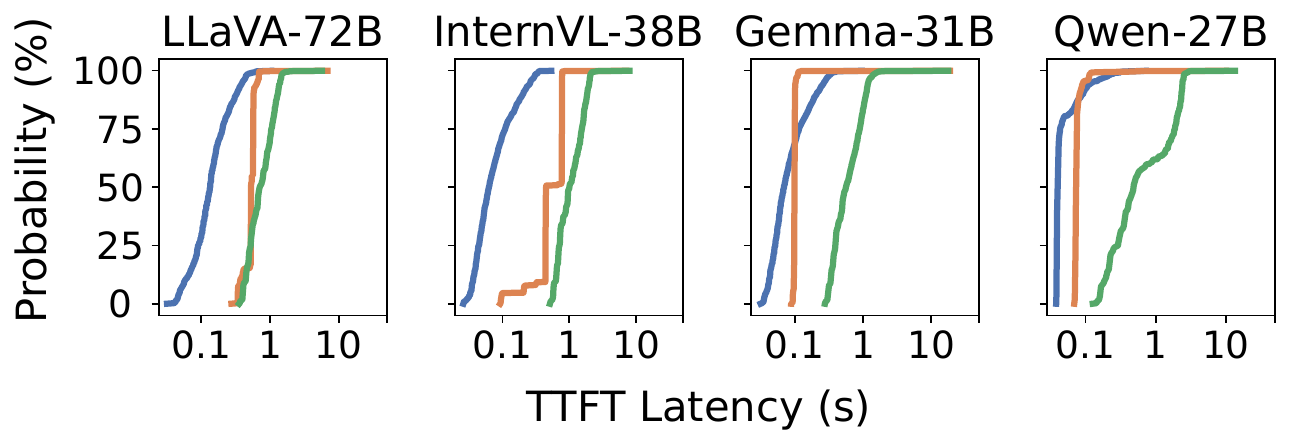}
        \caption{TTFT Latency}
        \label{fig:ttft-cdf}
    \end{subfigure}

    \caption{Characterization of different families of MLLMs.}
    \label{fig:mllm-char}
    \vspace{-0.1in}
\end{figure}

Using the multimodal workload described in Section~\ref{ssec:experimental-setup}, we run requests in isolation to characterize the impact of each modality type (text, image, and video) on performance and resources.
Figure~\ref{fig:mllm-char} reports the cumulative distribution of the KV Cache memory footprint, measured as the number of cached tokens, and time-to-first-token (TTFT) across multimodal LLMs. 

Figure~\ref{fig:mem-cdf} shows that memory footprint differs drastically across modalities, spanning multiple orders of magnitude across all models.
Text requests are the most lightweight, but also the most diverse, ranging from roughly $10$ up to $10^{4}$ cached tokens.
Image requests are consistently heavier, clustering tightly around $10^{3}$--$10^{4}$ tokens.
The near-vertical line for image requests reflects the fixed tokenization strategy of vision encoders: most models resize images and partition them into patches or tiles with standardized dimensions, producing an almost constant number of visual tokens per request.
Video requests are the most memory-intensive, often reaching $10^{4}$--$10^{5}$ cached tokens or more, as they contribute multiple image frames whose number grows with video length. Interestingly, there are overlaps across modalities.
For example, long text prompts may approach the footprint of image or video requests, and short videos may overlap with images.

\begin{figure}[t]
    \centering
    \includegraphics[width=\linewidth]{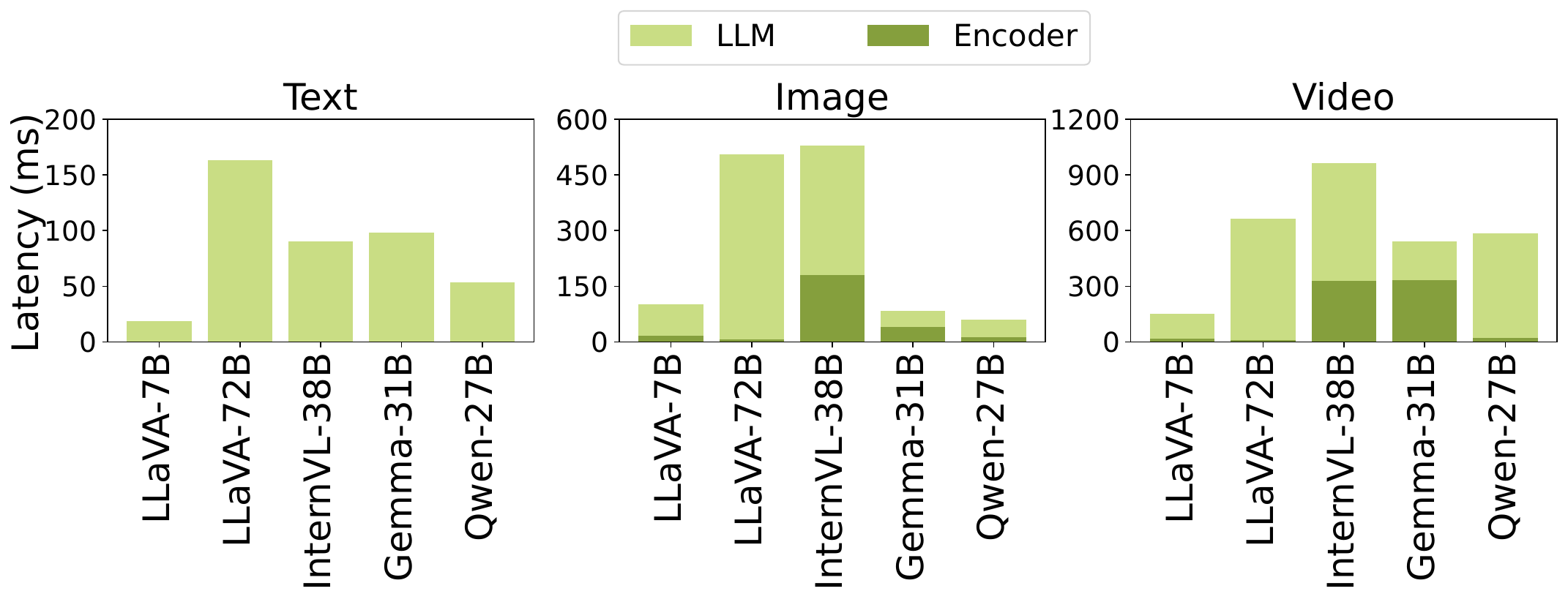}
    \caption{Breakdown of the GPU prefill latency.}
    \label{fig:ttft-breakdown}
    \vspace{-0.2in}
\end{figure}

Figure~\ref{fig:ttft-cdf} shows a similar pattern for latency, both in the separation across modalities and in the small overlaps between them.
Text requests are consistently the fastest, typically falling in the $0.01$--$0.1$ second range, although they still exhibit noticeable variance due to differences in prompt length.
Image requests are moderately slower, but generally complete in under $1$ seconds. Video requests are the most time-consuming, with TTFT typically ranging from $1$ to $10$ seconds.
Notably, both image and video requests exhibit a long TTFT tail, indicating that a subset of visual requests incurs substantially higher latency.
This behavior is consistent with observations from production-scale multimodal workloads~\cite{modserve,servegen}.

Figure~\ref{fig:ttft-breakdown} further decomposes TTFT into vision encoding and LLM prefill across modalities (text, image, video) and models. 
We observe substantial variation across model- modality pairs, highlighting the heterogeneous nature of multimodal inference. 
For example, models such as InternVL-38B and Gemma-31B exhibit significant vision encoding overhead, although their overall behavior differs: InternVL-38B incurs both high encoding and high LLM prefill latency, whereas Gemma-31B is primarily dominated by encoding, with comparatively lower LLM prefill cost.
In contrast, for large models such as LLaVA-72B and Qwen-27B, encoding overhead is negligible: LLM prefill dominates overall latency across modalities and, as expected, grows with model size.

\takeaways{
Multimodal requests differ by orders of magnitude in both \emph{space} (memory footprint) and \emph{time} (latency): videos are the most demanding, images are moderate, and text is extremely lightweight.
Encoding overheads depend on the model type.
These characteristics are consistent across multimodal models and align with production-scale multimodal workload characterizations~\cite{modserve,servegen}.
}

\subsection{Limitations of Current Approaches}
\label{ssec:limitations-of-sota}

The heterogeneous, bursty, and heavy-tailed characteristics of multimodal workloads~\cite{modserve, servegen}, together with their fundamentally different resource and latency profiles, have motivated a new class of serving systems and optimizations tailored to multimodal LLMs~\cite{modserve, elasticmm, epd}.
A key challenge is \emph{head-of-line (HoL) blocking}: multimodal requests, particularly image and video inputs, incur significantly longer and more variable processing times than text requests.
When co-scheduled, these expensive requests can delay lightweight text queries and degrade responsiveness, making it difficult to satisfy their TTFT latency objectives~\cite{gpt4-sla, dist-serve, dynamollm}.

A similar challenge has previously emerged in LLM serving, where requests with long input contexts delay shorter requests during prompt processing~\cite{longbench,sarathi-serve,sarathi,loongserve,mnemosyne}.
To mitigate this interference, among other approaches (Section~\ref{sec:related-work}), prior systems introduce \emph{disaggregation}, separating prefill and decode execution across distinct GPUs or nodes.
Systems such as DistServe~\cite{dist-serve}, Splitwise~\cite{splitwise} and Mooncake~\cite{mooncake} demonstrate that isolating long-running prefill operations can substantially reduce HoL blocking and improve serving performance.
Multimodal workloads further exacerbate this challenge with their latency and memory requirements varying by orders of magnitude across modalities (Figure~\ref{fig:mllm-char}).

Recent multimodal serving systems extend the disaggregation principle to account for the additional vision encoding stage~\cite{modserve, elasticmm, epd}. ModServe~\cite{modserve} separates the vision encoder from the LLM backend, isolating expensive encoding operations from text processing.
EPD~\cite{epd} further decomposes execution into independent encoding, prefill, and decode stages, enabling stage-specific optimizations.
ElasticMM~\cite{elasticmm} generalizes this approach through hierarchical disaggregation, separating workloads both by modality and by execution stage while dynamically adjusting resource allocation and parallelism.
Collectively, these systems reduce cross-stage interference by dedicating resources to the individual stages of the multimodal inference pipeline.

\custompar{Limitation}
While disaggregation effectively reduces interference, it introduces a fundamental {\it trade-off} between isolation and resource utilization: resources dedicated to one stage cannot be readily utilized by another.
Existing multimodal serving systems are primarily designed and evaluated in large-scale deployments (e.g., up to 128 GPUs~\cite{modserve}), where abundant GPU resources allow individual pipeline stages to be assigned dedicated hardware with minimal impact on overall utilization and performance.
However, many practical deployments operate at more modest scales, where the available GPUs must simultaneously provide high utilization and low latency.
In these settings, dedicating GPUs to specific stages can lead to resource fragmentation, raising the question of whether disaggregation remains the most effective design choice in this context.

To better understand this trade-off, we evaluate the performance and resource efficiency of disaggregation-based multimodal serving over 4 NVIDIA H100 GPUs.
We compare a monolithic vLLM~\cite{vllm} deployment, where MLLMs are configured with tensor parallelism degree 4 (TP4), against ModServe~\cite{modserve}, a representative disaggregation-based design.
Under a 4-GPU budget, ModServe can execute with a 2+2 partitioning, where 2 GPUs are dedicated to the vision encoder and the remaining 2 GPUs to the LLM backend (prefill and decode), each operating with TP2.
We use the NVIDIA API~\cite{nvidia-nvml} to continuously monitor GPU utilization, and report the time-averaged aggregate utilization across all 4 GPUs.

\begin{figure}[t]
    \centering
    \includegraphics[width=0.95\linewidth]{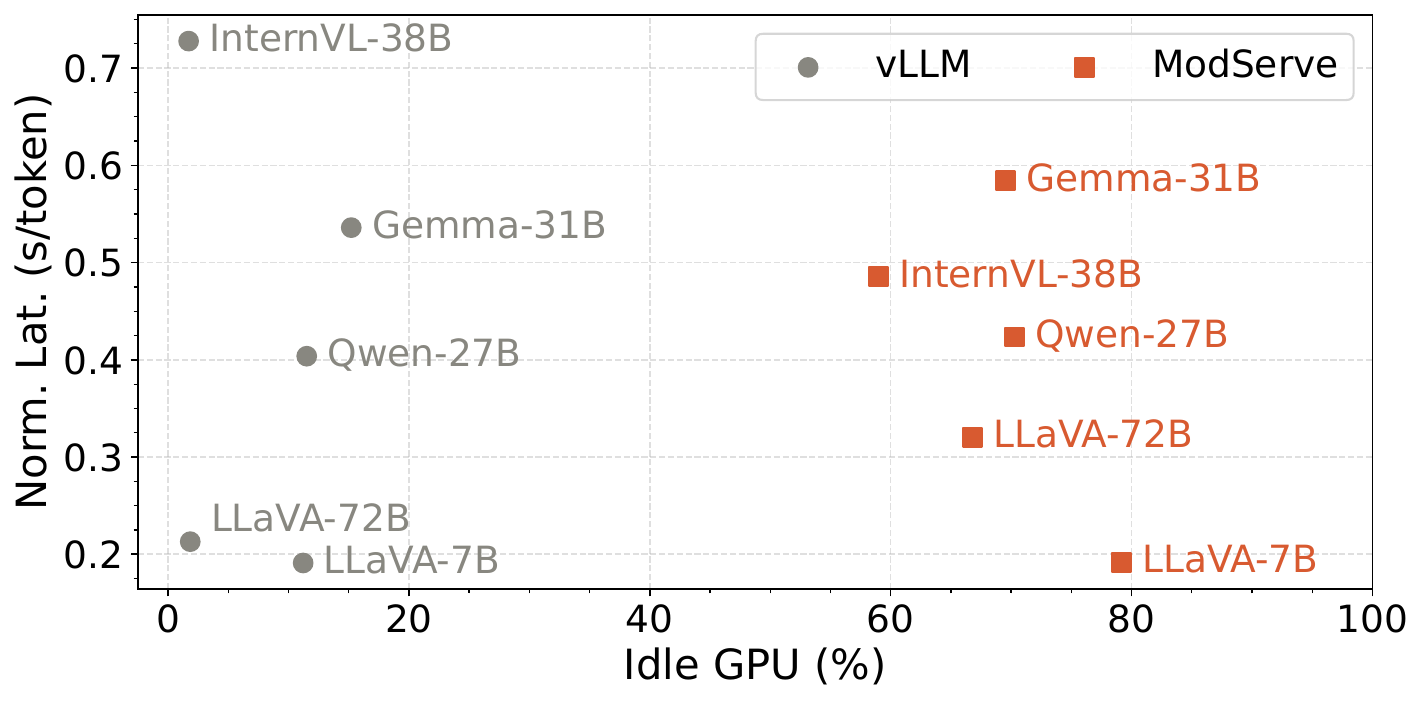}
    \caption{Trade-off between performance and GPU utilization of disaggregation-based multimodal serving systems.}
    \label{fig:limitations}
    \vspace{-0.2in}
\end{figure}

Figure~\ref{fig:limitations} illustrates the resulting trade-off between normalized latency (seconds / token) and resource utilization across MLLMs of varying sizes and families.
While disaggregation reduces interference by isolating the encoder from the LLM backend, it also limits resource sharing across stages.
In our setting, this leads to substantial underutilization, with approximately 60--70\% of aggregate GPU capacity remaining idle across all models.
The root cause is not the encoder itself, but the resource partitioning required to support stage isolation: resources allocated to the encoder cannot be dynamically repurposed when encoder demand is low.

In contrast, the monolithic vLLM deployment allows all 4 GPUs to be jointly utilized via TP4, reducing idle GPU capacity to approximately 1--2\% for LLaVA-72B and InternVL-38B, and 15--18\% for Qwen-27B and Gemma-31B. 
However, higher resource utilization does not consistently translate into better performance; rather, it depends on the model itself.
More specifically, it depends on the relative contributions of encoding and LLM prefill, as shown in Figure~\ref{fig:ttft-breakdown}.
For example, InternVL-38B exhibits around 50\% higher normalized latency under vLLM compared to ModServe, as encoding dominates its TTFT and disaggregation alleviates this bottleneck.
In contrast, LLaVA-72B achieves around 50\% lower normalized latency with vLLM compared to ModServe, since LLM prefill dominates and encoding overhead is negligible, making disaggregation unnecessary and even detrimental due to added fragmentation.
Finally, models such as Gemma-31B and Qwen-27B exhibit lower overall prefill costs, and vLLM still provides modest improvements of up to 10\% due to the improved utilization.

In conclusion, while vLLM significantly improves resource utilization, performance remains highly model-dependent: models with substantial encoding overhead benefit from disaggregation, whereas for models where encoding is negligible, disaggregation introduces unnecessary overheads and can degrade performance.

\takeaways{
The above analysis exposes a fundamental, {\it model-dependent trade-off} in recent multimodal serving systems at practical deployment scales. 
Disaggregation alleviates head-of-line blocking from vision encoding, but incurs poor GPU utilization due to partitioning limited resources across inference stages, and only benefits models where encoding dominates latency. 
In contrast, monolithic execution with FCFS scheduling achieves high resource utilization and can perform better for models where LLM prefill dominates, but remains vulnerable to severe head-of-line blocking under heterogeneous workloads. 
Together, these observations highlight the need for an effective multimodal serving solution that adapts to model-specific bottlenecks while jointly addressing heterogeneous request interference and resource efficiency, motivating the design of \system.
}
\section{System}
\label{sec:system}

\subsection{Overview and Objectives}
\label{ssec:system-overview}

This paper proposes \textbf{\system} (\textbf{R}ocks, \textbf{P}ebbles, and \textbf{S}and Serving), a modality-aware scheduling framework that explicitly accounts for the heterogeneous resource demands of multimodal inference workloads.
In multimodal workloads, resource and time requirements  differ by orders of magnitude across requests: some requests behave like {\it rocks}, dominating both latency and GPU memory; others resemble {\it pebbles}, incurring moderate overhead; while many requests are {\it sand}, lightweight but highly latency-sensitive.
The goal of \system is to {\it let the sand flow quickly through pebbles and rocks, maximizing GPU utilization}, while ensuring fast responsiveness for lightweight requests and controlled execution for heavier ones.

The design of \system is guided by two {\bf objectives}:
\begin{itemize}[leftmargin=25pt, itemsep=0pt, topsep=0pt]
    \item[\textbf{[O1]}] \textbf{High-performance multimodal inference}: Prioritizes latency-critical lightweight requests in return for controlled execution for heavier delay-tolerant ones.
    \item[\textbf{[O2]}] \textbf{Resource efficiency}: Maximize the utilization of the {\it available} GPU resources, avoiding the need for additional hardware to disaggregate inference stages and alleviate head-of-line (HOL) blocking.
\end{itemize}

\begin{figure}[t]
    \centering
    \includegraphics[width=0.9125\linewidth]{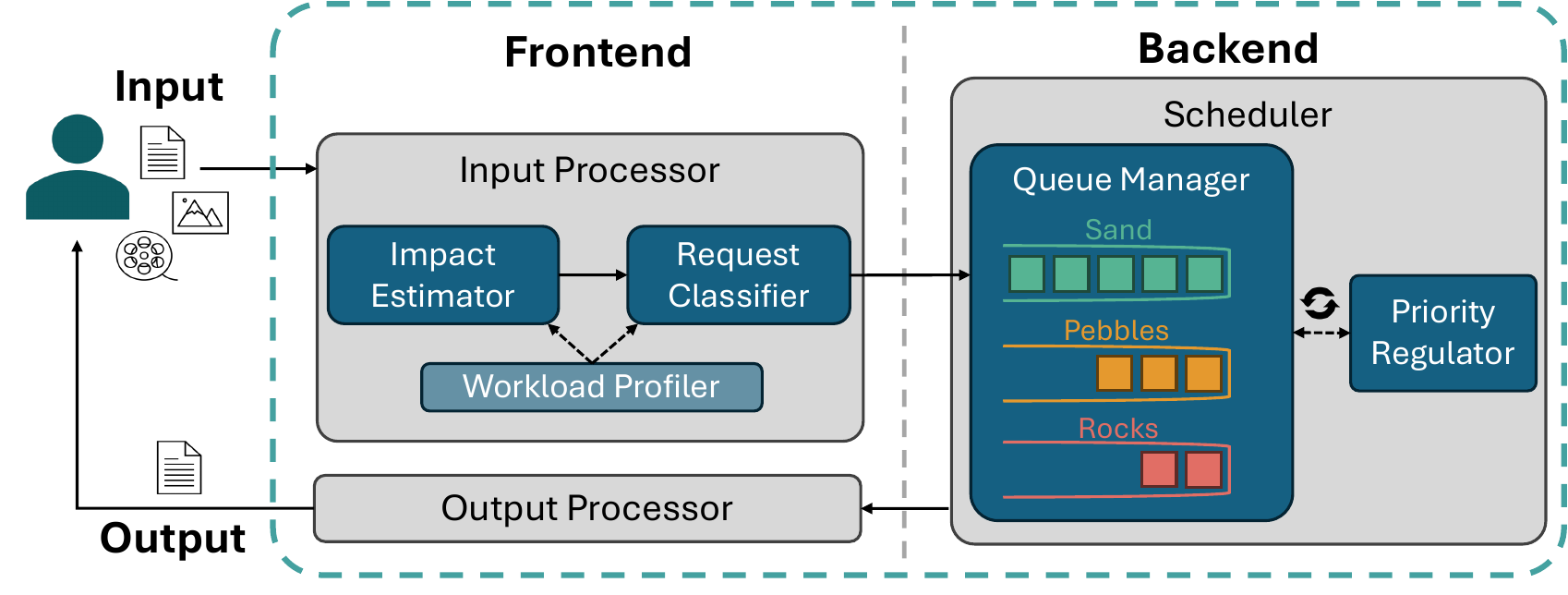}
    \caption{\system~System Components.}
    \label{fig:system-design}
    \vspace{-0.2in}
\end{figure}

To achieve these objectives, \system begins by estimating the GPU prefill latency and memory of each incoming request using model- and modality-specific estimators.
Prefill dominates batch latency~\cite{sarathi,sarathi-serve,loongserve, modserve, servegen}, making it critical for scheduling decisions.
The prefill (latency, memory) features capture the request's \emph{admission cost}.
Then, \system uses these features, instead of relying on modality alone (video, image, or text), to classify requests into three categories (rocks, pebbles, and sand) with lightweight unsupervised clustering.
\system maintains separate queues per category and places incoming requests at the tail of each corresponding queue.

\system performs priority-based scheduling, assigning higher priority to sand, followed by pebbles and rocks.
However, it is not limited to this static prioritization: \system incorporates waiting time to ensure that long-waiting requests are eventually scheduled across all classes.
At each scheduling iteration, \system updates the priority of all requests based on their class, waiting time, and queue state, which is reflected in the resulting ordering within each queue.
Then, the scheduler forms the next execution batch by selecting requests from the head of each queue following the global priority order across sand, pebbles, and rocks.
Depending on their priority, requests may be admitted into the batch (for prefill or to resume decoding), kept waiting, or preempted to make room for higher-priority requests.
This allows \system to dynamically adapt scheduling to the workload, reflecting the mix of latency-critical requests and the accumulated waiting time of heavier ones.

Figure~\ref{fig:system-design} illustrates the system components that implement the design of \system.

\noindent\textbf{Input Processing (Frontend):}
\begin{tightitemize}
    \item \textbf{Workload Profiler:} Builds offline performance profiles for multimodal models across text, image, and video of varying sizes (Section~\ref{ssec:workload-profiler}).
    \item \textbf{Impact Estimator:} Estimates the GPU admission cost of an incoming request by capturing its GPU prefill latency and memory footprint (Section~\ref{ssec:estimator}).
    \item \textbf{Request Classifier:} Classifies requests into rocks, pebbles, or sand based on (latency, memory) features using lightweight unsupervised clustering (Section~\ref{ssec:request-classifier}).
\end{tightitemize}

\noindent \textbf{Request Scheduling (Backend)}:
\begin{tightitemize}
    \item \textbf{Priority Regulator:} At each scheduling iteration, updates request priorities based on category, waiting time, and queue state, assigning higher priority to sand and long-waiting requests (Section~\ref{ssec:priority-regulator}).
    \item \textbf{Queue Manager:} Maintains separate queues for rocks, pebbles, and sand, tracks queue-level metrics, and enforces the priority ordering (Section~\ref{ssec:queue-manager}). 
    \item \textbf{Scheduler:} Employs a priority-based policy, following the global priority ordering across queues, and selects which requests will be part of the execution batch at each iteration (Section~\ref{ssec:scheduler}). 
\end{tightitemize}

\noindent
Together, these components instantiate the modality-aware scheduling principles of \system.
Modality-awareness is introduced through model- and modality-specific admission-cost estimation, which classifies requests as rocks, pebbles, or sand.
The backend then schedules these classes using priority-aware policies.
Although generally applicable to heterogeneous workloads, these policies are guided by modality-aware classification and thereby capture the distinct characteristics of text, image, and video requests.
Next, the functionality of each component is described in detail.

\subsection{Workload Profiler}
\label{ssec:workload-profiler}

The \textit{Workload Profiler} is an offline component that builds performance profiles for each model–modality pair, capturing modality-specific characteristics that strongly influence inference latency and memory footprint, as observed in Figures~\ref{fig:mem-cdf} and \ref{fig:ttft-cdf}.
To construct a profile, the system executes a representative workload for the target modality against the chosen MLLM, processing one request at a time to eliminate interference.
More specifically, it runs inference against the target MLLM for increasing size of text prompt lengths, images and videos using publicly available datasets, as the ones used throughout our experiments.
For each request, the profiler records the following key metrics: the time spent in the GPU for vision encoding and LLM prefill along with the number of tokens generated for the corresponding inputs.
Profiling is performed once per model during registration.
All collected data is stored as it will be later used by the {\it Impact Estimator} and the {\it Request Classifier}.

\subsection{Impact Estimator}
\label{ssec:estimator}

The \textit{Impact Estimator} captures the GPU admission cost of an incoming request, defined by its GPU prefill latency and KV Cache memory footprint.
Prefill latency refers to the time spent on the GPU during the prefill stage, including vision encoding and LLM prompt computation, and excludes CPU-side preprocessing (Figure~\ref{fig:mllm-arch})

Memory is calculated as the number of prefill tokens, used as a proxy for the KV Cache footprint, following prior work~\cite{orca,vllm,sarathi,konpap-euromlsys26}, as KV Cache size scales linearly with the number of tokens.
Latency is estimated using lightweight predictive models.
Figure~\ref{fig:ttft-breakdown} shows the breakdown of GPU prefill latency across modalities (text, image, video) and models of different families and sizes.
We observe substantial variation across model–modality pairs.
For example, models such as InternVL-38B and Gemma-31B spend a significant portion of time in vision encoding for image and video inputs. 

\begin{figure}[t]
    \centering
    \includegraphics[width=\linewidth]{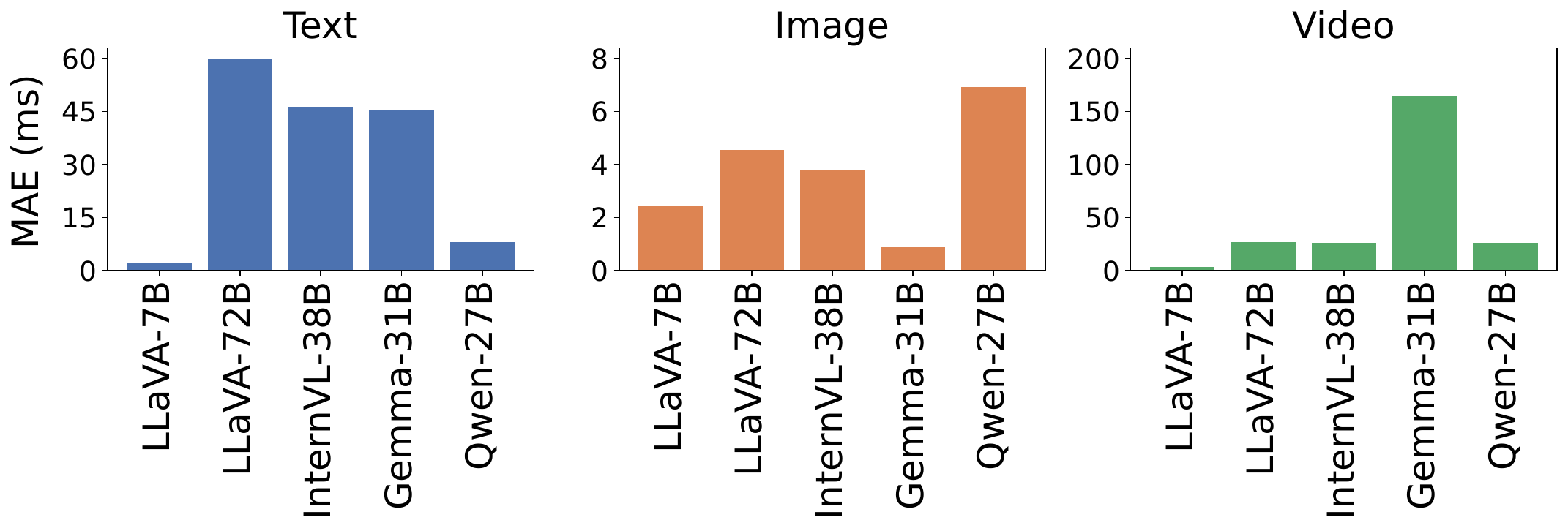}
    \caption{Estimation error of the GPU prefill latency.}
    \label{fig:estimator-error}
    \vspace{-0.2in}
\end{figure}

These differences motivate the need for model- and modality-aware latency estimators.
For text-only serving, prompt length is often a good proxy for request cost, motivating lightweight estimators used in prior LLM schedulers~\cite{sola, sloawaresched, intel-router, tempo}.
We therefore use a lightweight linear regression model, since prefill latency scales predictably with prompt length (token count).
In multimodal serving, however, admission cost depends on both model and modality due to vision encoding, whose contribution varies significantly across MLLMs, as shown in Figure~\ref{fig:ttft-breakdown}.
Consequently, request cost cannot be inferred from prompt length alone, motivating model- and modality-aware admission cost estimation.
Therefore, for image and video requests, we employ quantile regression (90\textsuperscript{th} percentile) to avoid potential underestimation.
Figure~\ref{fig:estimator-error} validates this design: prediction errors (mean absolute error in milliseconds) remain small relative to the corresponding GPU prefill times shown in Figure~\ref{fig:ttft-breakdown}.
This demonstrates that simple, model- and modality-specific estimators are sufficient to estimate prefill latency across modalities.
These estimators are trained offline and stored during system initialization using profiling data from the \textit{Workload Profiler}.
At runtime, the \textit{Impact Estimator} forwards the estimated latency and calculated memory to the \textit{Request Classifier}. 

\subsection{Request Classifier}
\label{ssec:request-classifier}

The \textit{Request Classifier} is central to \system because it realizes the rocks, pebbles and sand abstraction.
A naive modality-based classification (text $\rightarrow$ sand, image $\rightarrow$ pebbles, video $\rightarrow$ rocks) is insufficient because, in practice, latency and memory demands can overlap across modalities.
For example, long text prompts can be comparable to image or video requests, while short video requests can overlap with image requests, depending on the model, as shown in Figure~\ref{fig:mllm-char}.
Instead, \system classifies requests based on their GPU prefill (latency, memory) estimates.
These are provided by the \emph{Impact Estimator}, which uses model- and modality-specific predictors corresponding to the request's target MLLM and input modality (text, image, video).
This approach more accurately reflects the temporal and spatial footprint of each request. 
In this way, classification is {\it indirectly} modality-aware, since modality information influences the estimated admission cost used for classification.

For classification, the \textit{Request Classifier} employs a lightweight k-means clustering model.
Offline, it applies k-means (with Euclidean distance) to standard scaled latency and memory features collected by the \emph{Workload Profiler}, grouping requests into \(k = 3\) clusters. 
\system is designed around three request classes, namely \emph{sand}, \emph{pebbles}, and \emph{rocks}, which operationalize the Rocks-Pebbles-Sand (RPS) abstraction and enable differentiated scheduling policies for each class.
Using \(k=3\) provides a natural mapping between lightweight, moderate, and resource-intensive requests, capturing the dominant behaviors observed in multimodal workloads.
At the same time, it allows for a practical deployment, as each class comes with specialized scheduling policies and parameters.
Section~\ref{sec:sensitivity-analysis} further discusses this trade-off.
To label the clusters as \emph{sand}, \emph{pebbles}, or \emph{rocks}, \system computes the sum of latency and memory for each centroid and assigns the labels in increasing order of this score. 
Online, incoming requests are assigned to the nearest centroid using Euclidean distance, thereby determining their class.

\begin{figure}[t]
    \centering
    \includegraphics[width=\linewidth]{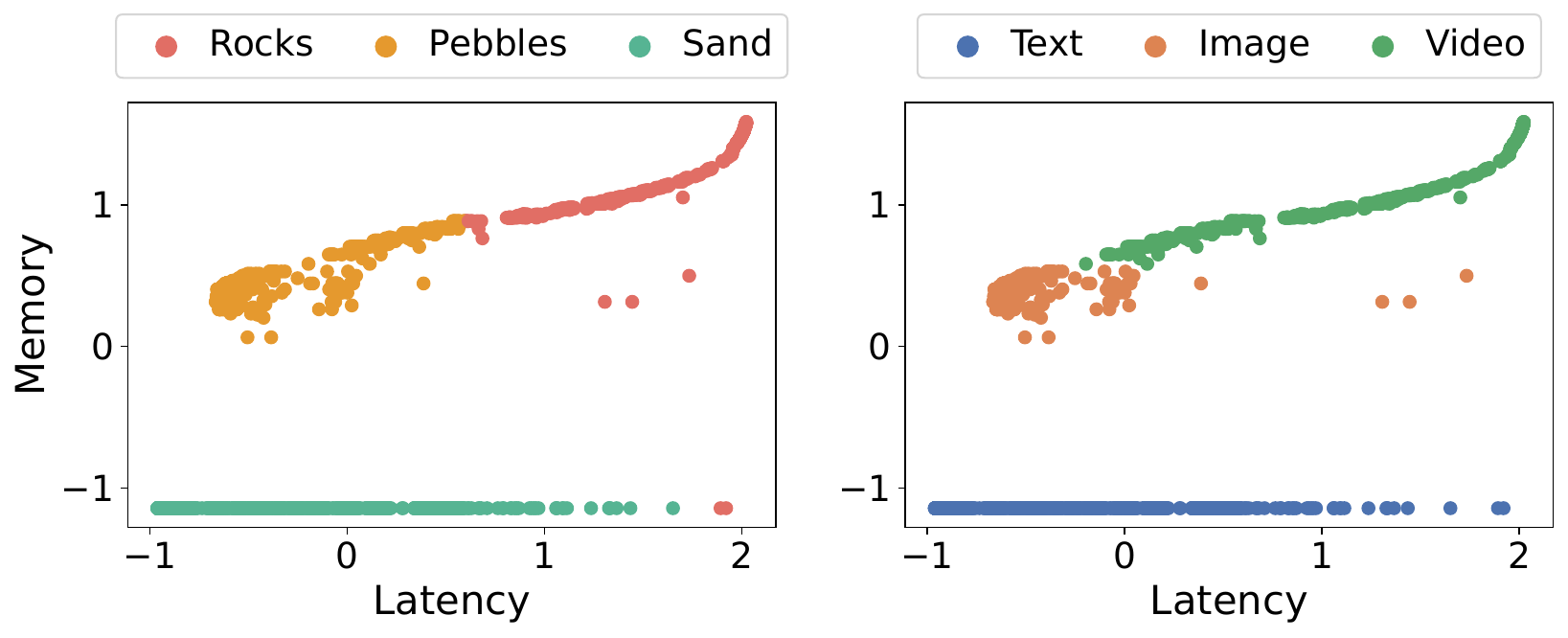}
    \caption{Comparison of request clustering based on \system's classifier (left) vs. on modality-alone (right).}
    \label{fig:clustering}
\end{figure}

Figure~\ref{fig:clustering} compares \system’s clustering-based classification (left) with a naive modality-based grouping (right) when using Qwen-27B.
Each point represents a request in a projected 2D latency-memory space.
We observe that \system groups requests based on their modality-aware admission cost rather than just their modality, resulting in well-separated clusters that reflect actual system impact.
In contrast, the modality-based grouping mixes requests with significantly different latency and memory characteristics.
Notably, some short video requests fall into the pebble cluster, while some long text requests are classified as rocks, highlighting that modality alone is not sufficient. 
Section~\ref{ssec:ablation} further evaluates the impact of classification to inference performance and Section~\ref{sec:sensitivity-analysis} the impact of the number of clusters.

\subsection{Priority Regulator}
\label{ssec:priority-regulator}

The \emph{Priority Regulator} dynamically updates the priority of each request at every scheduling iteration.
Upon request arrival, it assigns a $\text{StaticPriority}_c$ based on the classification $c$ provided by the \emph{Request Classifier}, mapping sand, pebbles, and rocks to decreasing priority tiers.
However, such a static prioritization can lead to imbalanced performance across request classes: lightweight sand requests may still experience GPU resource contention and delays, while heavier pebble and rock requests risk excessive scheduling delays.
To further improve responsiveness and fairness, \system augments the $\text{StaticPriority}_c$ with an aging term that increases with the time a request spends waiting in the scheduling queue. The priority of a request \(r\) at scheduling iteration \(i\) is: 
\[
\begin{gathered}
\text{Priority}_r(i) = \text{StaticPriority}_c + \left(1 - e^{-k_c(i) \cdot (\text{waiting\_time}_r(i)^{p_c})}\right),\\
\text{waiting\_time}_r(i) = t_i - t_{\text{arrival}}(r).
\end{gathered}
\]
where \(c \in \{\text{Sand}, \text{Pebbles}, \text{Rocks}\}\).
Since the aging term is a monotonic function of waiting time, which is determined by arrival order, the policy implicitly incorporates FCFS per class: earlier requests accumulate priority over time and are eventually scheduled.
The term \(k_c(i)\) captures the queue pressure, that is the load, the length and average waiting time for class \(c\) at iteration \(i\), causing requests in longer and slower queues to gain priority more rapidly.
The term \(p_c\) statically controls how quickly priority grows with waiting time, with higher values for sand, followed by pebbles and rocks.
Consequently, the aging term increases fastest for sand, more gradually for pebbles, and slowest for rocks.

This design reflects the inherent latency asymmetry across modalities in multimodal workloads (Figure~\ref{fig:ttft-cdf}): text requests require interactive response times (sub-second), whereas image- and video-based requests are naturally more delay-tolerant~\cite{modserve, servegen}, often incurring delays of several to tens of seconds before response generation begins.
Thus, allowing controlled additional waiting time for heavier requests aligns with their expected execution characteristics while preserving responsiveness for latency-critical workloads.
Section~\ref{ssec:baseline-compare} evaluates the impact of this design approach on performance and SLO adherence.

\subsection{Queue Manager}
\label{ssec:queue-manager}

The \textit{Queue Manager} maintains three separate queues, one per request class \(c \in \{\text{Sand}, \text{Pebbles}, \text{Rocks}\}\).
At each scheduling iteration \(i\), it interacts with the \emph{Priority Regulator} to determine the updated ordering of requests.
Incoming requests are appended to the tail of their corresponding class queue and initialized with \(\text{Priority}_r(i) = \text{StaticPriority}_c\), since their waiting time is zero.
At every iteration, the \textit{Queue Manager} tracks queue-level statistics, including the queue length \(k_c(i)\) and the waiting time \(\text{waiting\_time}_r(i) = t_i - t_{\text{arrival}}(r)\) for each request.
These statistics are passed to the \emph{Priority Regulator}, which computes the updated priorities and returns them to the \textit{Queue Manager} to enforce the priority ordering across queues.
Separating requests into class-specific queues decouples classification from scheduling, enabling class-specific scheduling without cross-class interference.

\subsection{Scheduler}
\label{ssec:scheduler}

The \textit{Scheduler} executes the priority-based policy and, at each iteration, dynamically determines the set of requests that will form the execution batch.
It selects from the heads of each class-specific queue as many requests as they fit in the batch size, following strict global priority ordering across sand, pebbles, and rocks.
Based on this ordering, requests may be admitted into the batch (e.g., for prefill or to resume decoding), deferred to future iterations, or preempted to make room for higher-priority requests.

\begin{figure}[t]
    \centering
    \includegraphics[width=\linewidth]{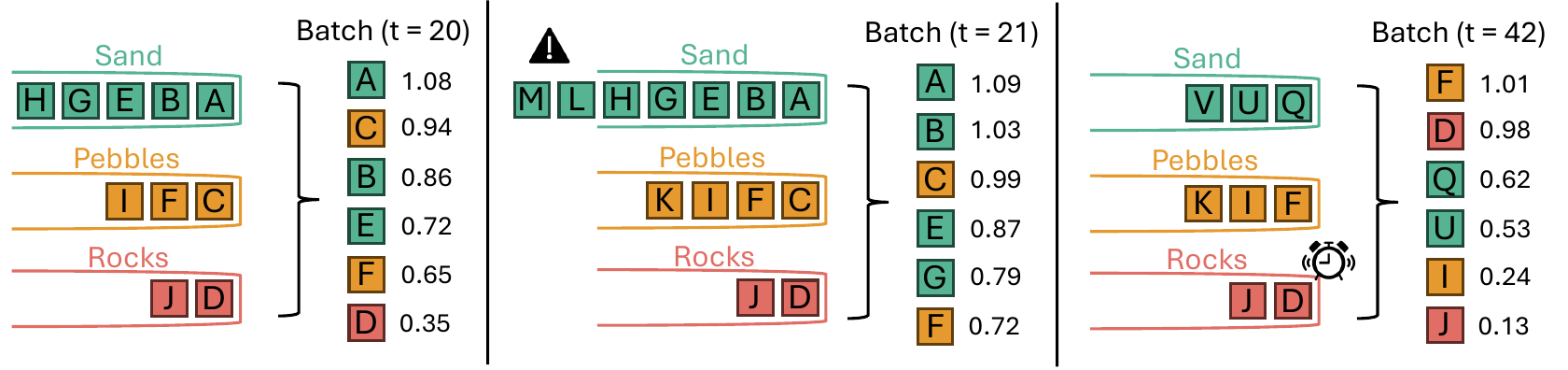}
    \caption{Example of batch composition across iterations.}
    \label{fig:batch-example}
    \vspace{-0.2in}
\end{figure}

Figure~\ref{fig:batch-example} illustrates how the scheduler dynamically adjusts batch composition across iterations based on the global priority ordering.
Requests are labeled alphabetically (A, B, C, \dots) in order of arrival, and are organized into class-specific queues (sand, pebbles, rocks).
Starting at an arbitrary iteration \(t{=}20\), the batch contains requests from all classes.
Notably, pebble request \(C\) has higher priority than some sand requests due to its longer waiting time.
Within each class, priorities follow arrival order, as all requests share the same static priority and differ only in their waiting time.
At \(t{=}21\), new sand requests (e.g., \(M\), \(L\)) arrive, increasing the sand queue length and pushing the urgency of existing sand requests.
As a result, the waiting sand request $G$ is scheduled, preempting the lower-priority rock request $D$.
By \(t{=}42\), the effect of aging becomes evident.
Requests that have been waiting for a long time, including pebbles and rocks, have accumulated sufficient priority to enter the batch.
In this way, \system balances low-latency responsiveness with continuous progress for all requests.

\section{Evaluation}
\label{sec:evaluation}

We evaluate \system to demonstrate its ability to meet its system design objectives, that is high-performance multimodal inference and maximizing the utilization of the available GPU resources.
Section~\ref{ssec:experimental-setup} describes in detail the experimental setup.
Section~\ref{ssec:baseline-compare} compares in-depth the performance of \system against the baselines and Section~\ref{ssec:ablation} discusses its ablation study and Section~\ref{sec:sensitivity-analysis} the sensitivity analysis.

\subsection{Experimental Setup}
\label{ssec:experimental-setup}

\custompar{Hardware} 
We experiment on a server with native hardware that includes 4 NVIDIA H100 GPU with 96GB memory each, two Intel® Xeon® Platinum 8592+ 64-Core processors (128 threads) and 1TB of host DRAM memory.

\custompar{Models} 
We experiment with state-of-the-art multimodal LLMs of different families and sizes, including LLaVA-7B~\cite{llava-ov,li2025llavaonevision} and LLaVA-72B~\cite{llava-ov-large,li2025llavaonevision}, InternVL3.5-38B~\cite{internvl-3.5-large,wang2025internvl3_5}, Gemma4-31B~\cite{gemma-4-large} and Qwen3.5-27B~\cite{qwen-3.5-large,qwen3.5}.

\custompar{Workloads} 
The evaluation workload captures the key characteristics reported by production-scale multimodal serving traces: burstiness, heavy-tailed request sizes, and variability both within and across modalities~\cite{modserve, servegen}, consistent with our characterization in Section~\ref{ssec:characterization-iso}.
As ModServe's~\cite{modserve} traces are not publicly available, and ServeGen's~\cite{servegen} includes text and image workloads but no video traces, we create a workload with 40\% short text, 30\% long text (>1K tokens), 20\% image, and 10\% video requests, drawn from ShareGPT~\cite{sharegpt} (text conversations), LLaVA-Instruct~\cite{liu2023llava} (image reasoning), and LLaVA-Video~\cite{llava-video-web} (video understanding).
Section~\ref{sec:sensitivity-analysis} further evaluates \system's sensitivity to workload composition.
Requests arrive through a single Poisson process shared across modalities. We intentionally do not model burstiness through the arrival process itself; instead, burstiness arises from the workload composition, where videos are treated as bursts of image frames~\cite{servegen}.

\begin{table}[t]
    \centering
    \small
    \begin{tabular}{c|c|c|c|c}
        \toprule
        \textbf{\#GPUs} & \textbf{vLLM} & \textbf{\system} & \textbf{ModServe} & \textbf{ModServe \& RPS} \\
        \midrule
        1 & TP1 & TP1 & \multicolumn{2}{|c}{--} \\
        2 & TP2 & TP2 & \multicolumn{2}{|c}{1E + 1P/D (TP1)} \\
        4 & TP4 & TP4 & \multicolumn{2}{|c}{2E + 2P/D (TP2)} \\
        \bottomrule
    \end{tabular}
    \caption{Baseline configurations under different GPU budgets. E: encoder. P/D: Prefill/Decode. TP$k$: tensor parallelism over $k$ GPUs.}
    \label{tab:config-gpu}
    \vspace{-0.2in}
\end{table}

\custompar{Baselines}
We evaluate \system against:
\begin{tightitemize}
    \item \textbf{vLLM~\cite{vllm}}: The state-of-the-art open-source LLM serving system, using default FCFS (first-come-first-served) scheduling.
    \item \textbf{ModServe~\cite{modserve}}: The state-of-the-art multimodal serving system that disaggregates the vision encoder from the LLM backend, similar to other recently proposed multimodal solutions, such as EPD~\cite{epd} and ElasticMM~\cite{elasticmm}.
    \item \textbf{ModServe \& RPS}: ModServe augmented with our modality-aware scheduling policy, isolating the impact of our proposed scheduling under a disaggregated deployment.
\end{tightitemize}

\custompar{Baseline Configuration Across GPUs}
Table~\ref{tab:config-gpu} summarizes the baseline configurations under different GPU budgets.
For single-node multi-GPU deployments, vLLM uses tensor parallelism (TP) as its standard distributed inference strategy~\cite{vllm_distributed_serving}.
With a single GPU, disaggregation-based baselines are not applicable.
For fair comparison, ModServe and ModServe \& RPS divide available GPUs evenly between the encoder and prefill/decode stages, while vLLM and \system use all GPUs for tensor parallelism.
We use the NVIDIA API~\cite{nvidia-nvml} to continuously monitor GPU utilization and report the time-averaged aggregate utilization across all GPUs.

\custompar{Implementation Details and Parameter Configuration}
\system is built on top of vLLM~\cite{vllm} (version 0.21.1, V1 engine).
We augment the scheduler~\cite{vllmSchedulerConfig} to support dynamic priorities and multiple queues.
vLLM uses scores to determine scheduling priority (lower score, higher priority).
We convert \system's priority of a request $r$ at iteration $i$ to \(\text{Score}_r(i) = -\log(\text{Priority}_r(i))\). 

The Priority Regulator of \system is configured with the following per-class parameters (StaticPriority, $p, k$): Sand: (0.1, 3.5, 0.05) Pebbles: (0.05, 2.5, 0.003) Rocks: (0, 1.1, 0.00075).
StaticPriority defines the class ordering (sand > pebbles > rocks), while the aging parameters $p, k$ control how quickly waiting requests gain priority.
Sand ages the fastest so that it gets scheduled the quickest, while rocks the opposite.
Since the aging term is bounded in [0,1) old pebble/rock requests can eventually overtake newly arrived sand requests and complete execution (Figure~\ref{fig:batch-example}).

\system inherits vLLM's recomputation-based preemption, which is invoked only when a higher-priority request cannot be admitted under the current resource constraints, minimizing unnecessary preemptions while preserving prioritization.
In addition, both \system and the vLLM baseline use vLLM's default chunked-prefill configuration~\cite{vllm_chunked_prefill}.
\system doesn't modify chunking.
It actually leverages chunk boundaries as scheduling opportunities, allowing, e.g., sand requests to interleave with long rock prefills.
Multimodal embeddings remain cached until prefill completes, as per the default vLLM behavior. 

We build the disaggregation-based baseline using vLLM's support for disaggregated encoder architecture~\cite{vllm_disagg_encoder}, since ModServe~\cite{modserve} is not open-sourced.
We configure ModServe’s \textit{Image Instances} as dedicated encoder nodes and its \textit{Text Instances} as joint prefill and decode nodes.
Then, we extend this implementation with \system's system components to support the ModServe \& RPS baseline.

\subsection{Comparison Against Baselines}
\label{ssec:baseline-compare}

\begin{figure}
    \includegraphics[width=\linewidth]{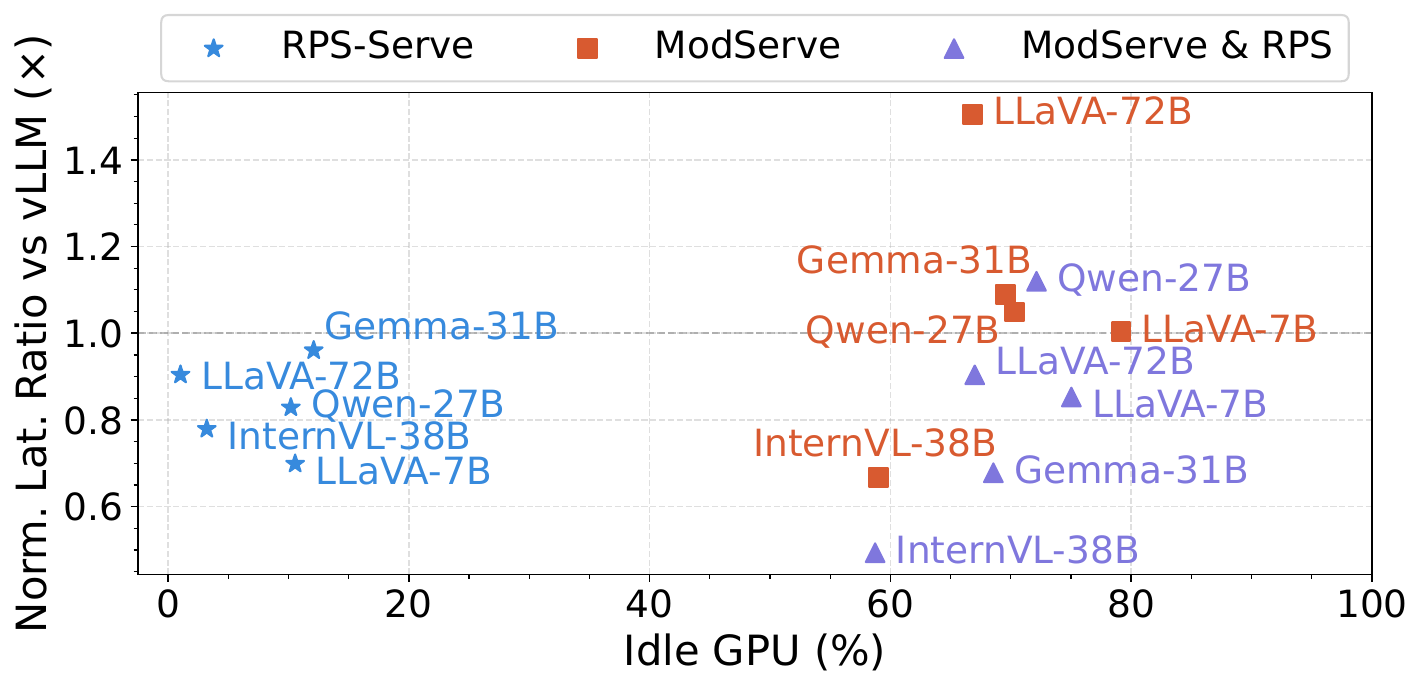}
    \caption{Trade-off between inference performance (relative to vLLM) and \% of idle GPU.}
    \label{fig:eval-perf-util}
\end{figure}

\customsection{Across Models} 
Figure~\ref{fig:eval-perf-util} shows normalized latency (seconds / token) as a function of idle GPU capacity across baselines and models in the 4-GPU setup, extending the trends observed in Figure~\ref{fig:limitations}.
The y-axis is normalized to vLLM (1$\times$), and the Pareto frontier corresponds to the bottom-left region.

We observe that \system consistently operates within the Pareto-optimal region.
Specifically, \system achieves 5--30\% (0.95--0.7$\times$) performance improvements over vLLM, while maintaining very low idle GPU capacity (1--15\%).
In contrast, ModServe operates in a clearly suboptimal regime, incurring 60--80\% idle GPU.
Its performance varies significantly across models: it is up to 50\% slower than vLLM for LLaVA-72B, slightly worse (0--10\%) for LLaVA-7B, Qwen-27B, and Gemma-31B, and only provides gains (up to 35\%) for InternVL-38B. 
This is due to the overheads of the encoding stage, which vary depending on the model family (Figure~\ref{fig:ttft-breakdown}).
For InternVL-38B, encoding dominates the GPU prefill latency, making disaggregation beneficial, whereas for LLaVA-72B, LLM prefill dominates, causing disaggregation to actually hurt performance.
Augmenting ModServe with \system (ModServe \& RPS) preserves similar GPU utilization, as it remains constrained by the underlying disaggregation, but substantially improves performance over ModServe alone: up to 40\% for LLaVA-72B and Gemma-31B and 25\% for InternVL-38B.

These results demonstrate that \system {\it successfully realizes its design objectives}: (O1) delivering high inference performance by reducing head-of-line blocking among heterogeneous requests, and (O2) maintaining high resource efficiency, as reflected by consistently higher GPU utilization under the same model and GPU budget, without requiring additional GPUs or disaggregation of inference stages.
Modality-aware, priority-based scheduling is the key enabler of these gains, providing consistent performance improvements across models and translating available GPU resources into useful work more effectively.
In contrast, disaggregating the encoder is only beneficial in cases where the encoding stage dominates latency (e.g., InternVL), and can otherwise degrade performance when LLM prefill is the primary bottleneck (e.g., LLaVA-72B). 
Across any type of models, combining disaggregation with modality-aware scheduling (ModServe \& RPS) further improves performance, highlighting that improved scheduling is the dominant factor for achieving better performance in multimodal serving.

\begin{figure}
    \centering
    \begin{subfigure}[b]{\linewidth}
      \includegraphics[width=\linewidth]{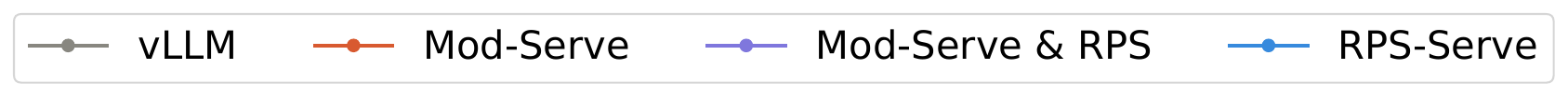}
    \end{subfigure}
    
    \begin{subfigure}[b]{\linewidth}
      \includegraphics[width=\linewidth]{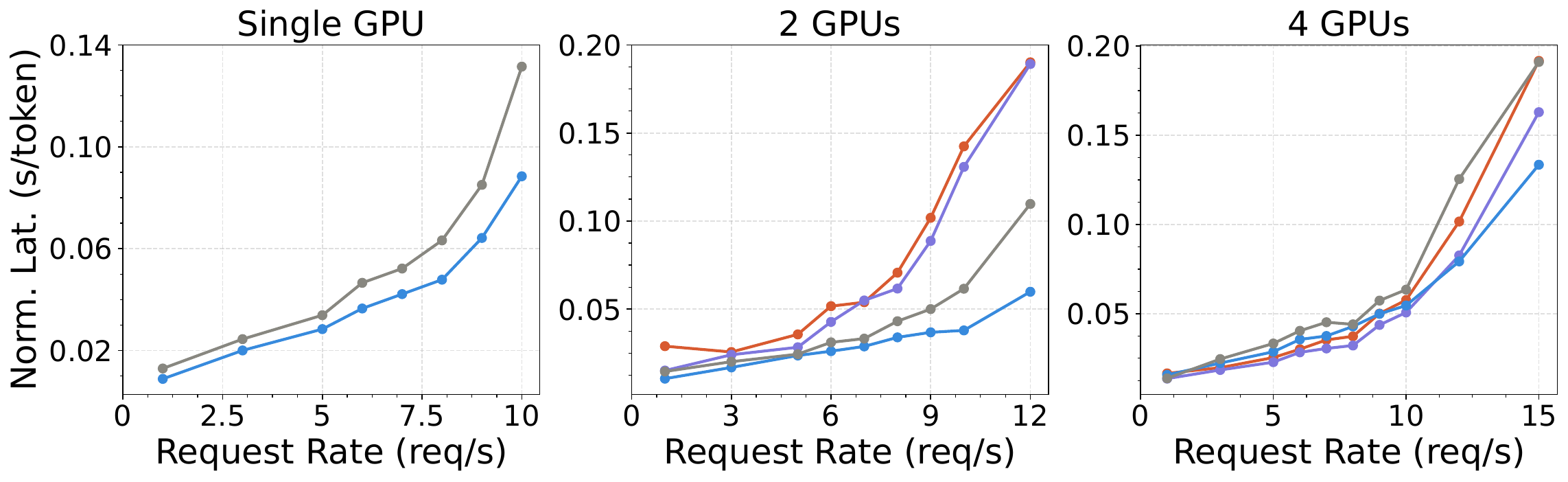}
      \caption{LLaVA-7B}
      \label{fig:latency-gpus-llava-small}
    \end{subfigure}
    
    \begin{subfigure}[b]{\linewidth}
      \includegraphics[width=\linewidth]{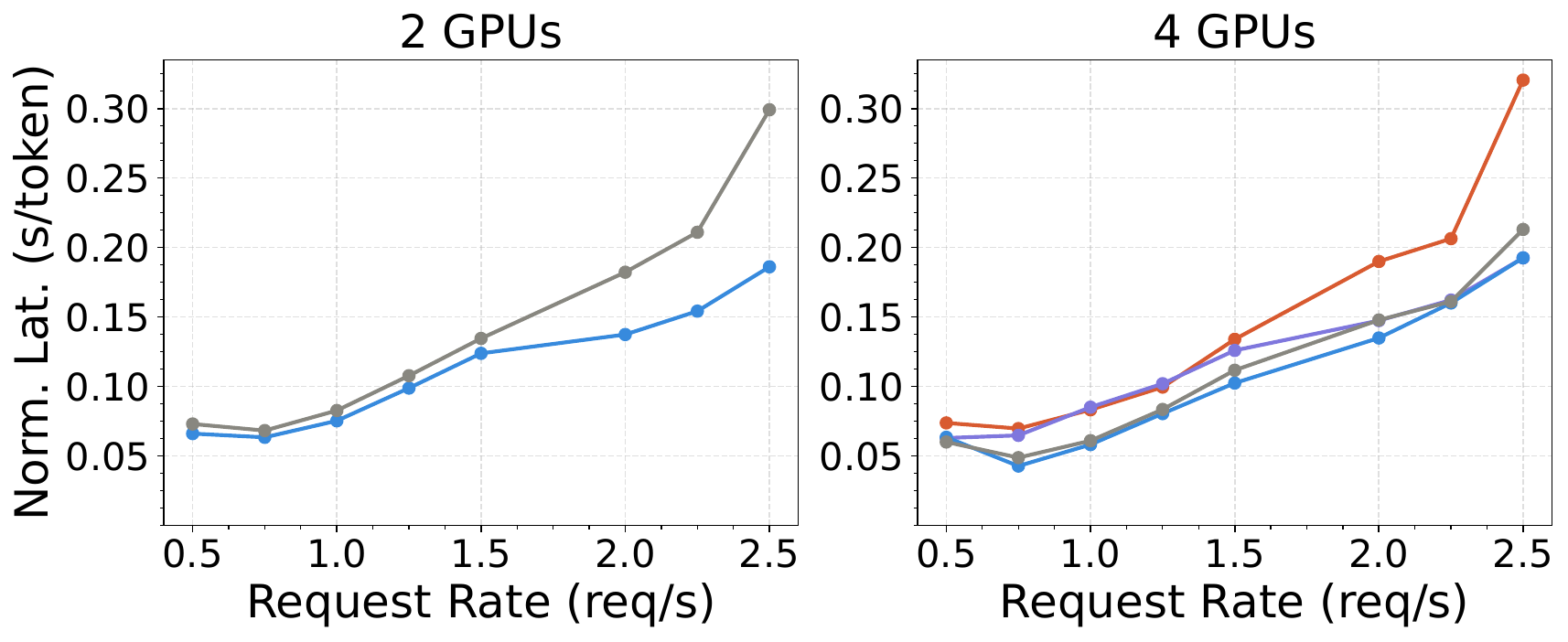}
      \caption{LLaVA-72B}
      \label{fig:latency-gpus-llava-large}
    \end{subfigure}
    
    \caption{Baseline performance comparison across increasing load and number of GPUs.}
    \label{fig:eval-load-gpus}
\end{figure}

\customsection{Under Increasing Load and Number of GPUs} 
Figure~\ref{fig:eval-load-gpus} shows normalized latency as a function of increasing load (request rate) across the baselines for different GPU budgets, using the configurations described in Table~\ref{tab:config-gpu}.
For LLaVA-7B, we evaluate configurations with 1, 2, and 4 GPUs, as it is the only model that fits on a single GPU, demonstrating that \system can operate and provide performance benefits even in the most constrained setting.
For LLaVA-7B, as the request rate increases and head-of-line (HoL) delays begin to dominate, \system delivers significant performance improvements up to 70\% lower normalized latency over all baselines regardless of the number of GPUs. 
For LLaVA-72B, because of its large size it requires at least 2 GPUs to run, so we evaluate all baselines in the 4-GPU setup.
For 2 GPUs it delivers 40\% better performance than vLLM under high load. For 4-GPU it is 40\% better than ModServe under high load, but the rest of the baselines are comparable as also shown in Figure~\ref{fig:eval-perf-util}, due to the large size of the model where LLM prefill dominates.
Overall, \system scales effectively with increasing load and GPU count, consistently maintaining high inference performance.

\begin{figure}[t]
    \centering
    \includegraphics[width=\linewidth]{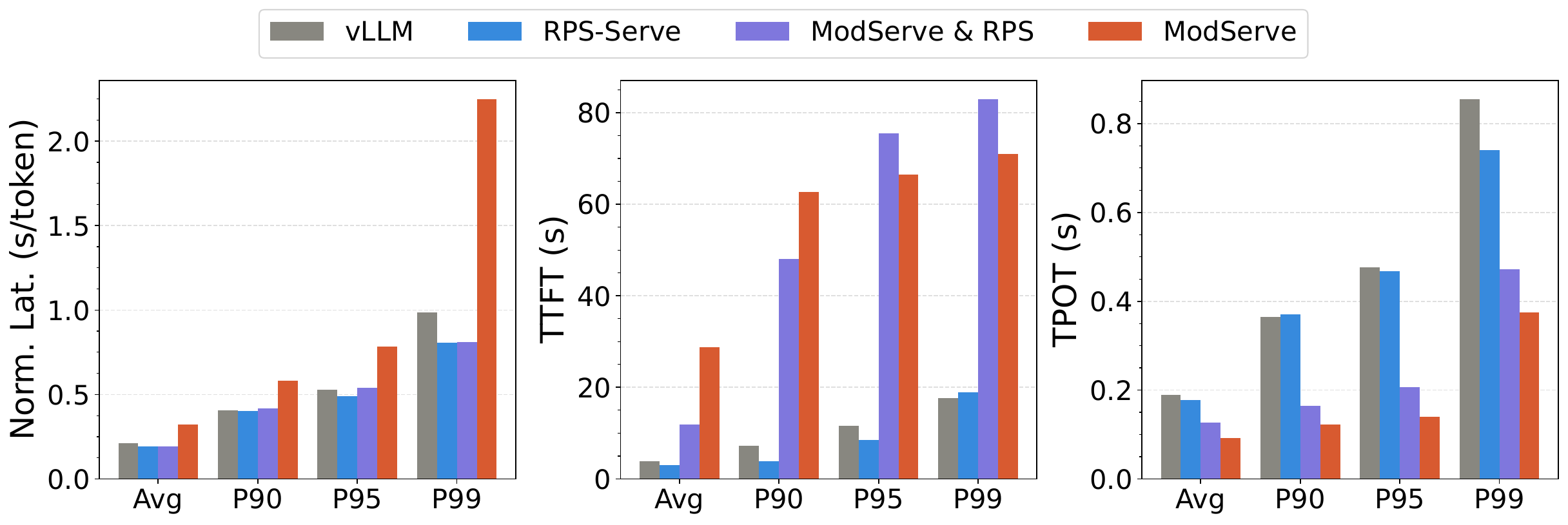}
    \caption{Baseline comparison of tail performance.}
    \label{fig:tail-breakdown}
\end{figure}

\customsection{Tail Performance}
Figure~\ref{fig:tail-breakdown} reports normalized latency, TTFT (time-to-first-token) and TPOT (time-per-output-token) for LLaVA-72B on 4 GPUs, comparing average-case and tail behavior (P90, P95, P99). 
Normalized latency follows the trends observed so far: \system, vLLM, and ModServe \& RPS achieve comparable performance even at the tail, while ModServe exhibits significantly higher normalized latency at P99, indicating that a small fraction of requests experience substantial slowdowns.
TTFT, however, reveals a more pronounced difference.
Compared to vLLM, \system reduces TTFT on average and at p90, P95, demonstrating that improved scheduling effectively mitigates head-of-line blocking and enables faster response initiation across modalities.
In contrast, both disaggregation-based baselines exhibit very poor tail behavior, with TTFT ranging from 48 up to 82 seconds at the tail, when vLLM and \system are in the order 3 to 18 seconds.
At the average case, \system reduces TTFT by 9.3$\times$ (28 down to 3 seconds) compared to ModServe. 
This reduction in TTFT comes with a trade-off in TPOT that is modest on average and becomes more substantial at the tail, especially compared to disaggregation-based baselines.
This occurs because lower-priority pebble and rock requests can get preempted to accommodate higher-priority sand requests, introducing additional delays in their output generation.

\begin{figure}[t]
    \centering
    \includegraphics[width=\linewidth]{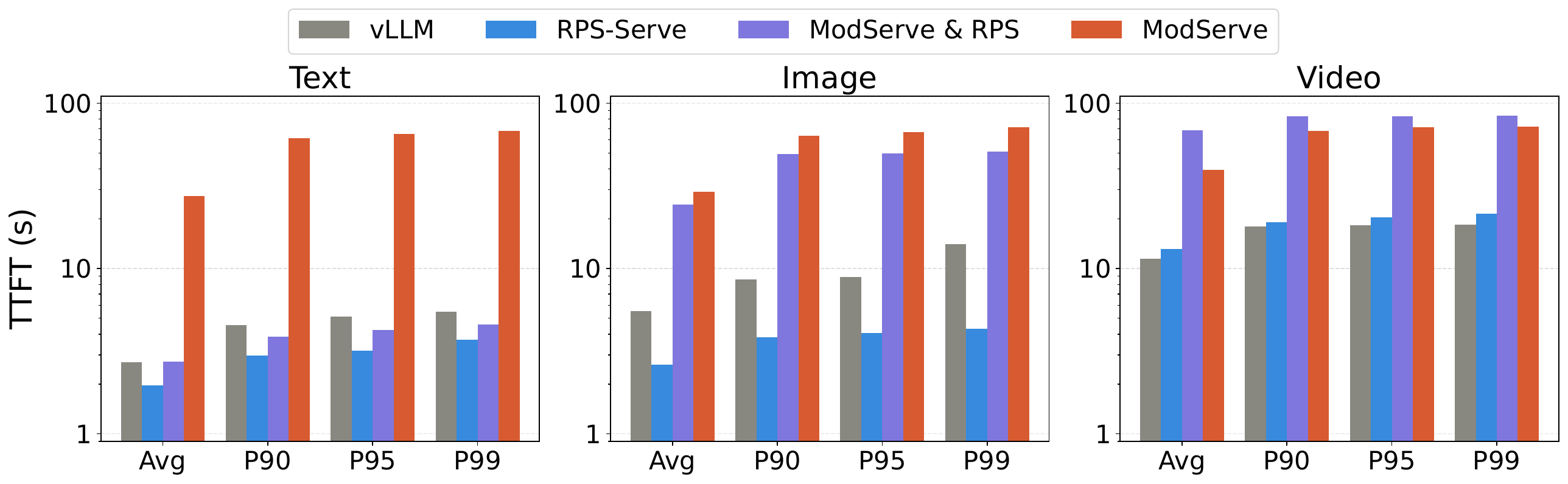}
    \caption{Baseline comparison of TTFT across heterogeneous modalities (text, image, and video).}
    \label{fig:perf-modalities}
\end{figure}

\customsection{Impact Across Modalities}
Figure~\ref{fig:perf-modalities} presents TTFT performance (in log scale) across modalities by comparing average-case and tail behavior (P90, P95, P99) for LLaVA-72B on 4 GPUs.
For text requests, \system achieves the lowest TTFT ranging from an average of 1.96 seconds (due to the very large model size) up to 3.7 seconds at the p99 tail, consistent with its design to prioritize and accelerate lightweight requests.
In contrast, ModServe incurs prohibitively high TTFT for text (more than 25 seconds).
Notably, in the case of ModServe \& RPS, it is \system that helps ModServe accelerate text to levels similar to vLLM, showing how request prioritization is the key to enable fast responsiveness for lightweight text requests. 

For image requests, \system reduces TTFT by 3--10 seconds compared to vLLM across average and tail, due to the decreased head-of-line blocking, while both ModServe and ModServe \& RPS remain significantly slower.
For video requests, \system incurs slightly higher TTFT than vLLM (by around 1.5--3 seconds), reflecting \system's deliberate trade-off that slightly delays resource-intensive requests.
In contrast, the disaggregation-based baselines exhibit extremely high TTFT for videos.
For both images and videos, adding \system to ModServe (ModServe \& RPS) does not bring performance benefits, because LLaVA-72B has minimal vision encoding overhead, therefore encoder disaggregation actually hurts performance, as seen before. 

Overall, \system prioritizes lightweight and medium-cost requests while only modestly delaying heavier ones, achieving strong performance both on average and at the tail, avoiding the extreme scheduling delays observed in the disaggregation-based baselines.

\begin{figure}[t]
    \centering
    \includegraphics[width=\linewidth]{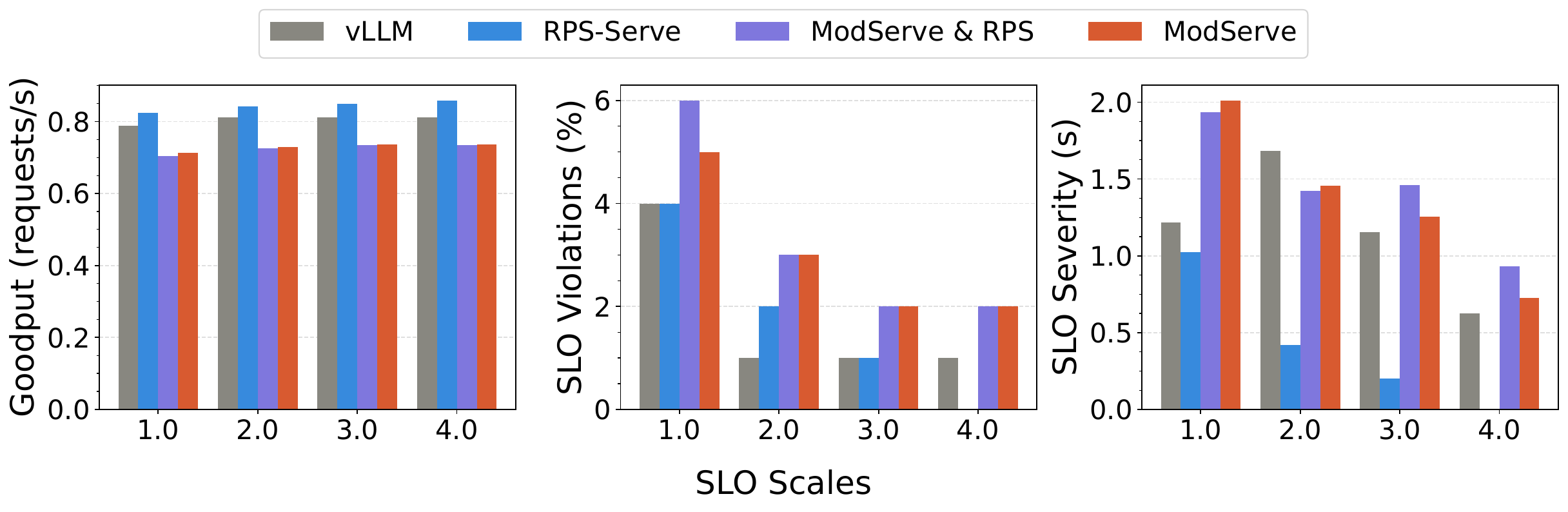}
    \caption{Baseline comparison of goodput, SLO violations and average violation severity under increasing SLO scales.}
    \label{fig:slo-modalities}
\end{figure}

\customsection{Goodput and SLO Violations}
Next, we compare baselines under explicit performance objectives.
Figure~\ref{fig:slo-modalities} reports goodput (requests per second), SLO (Service-level Objectives) violations, and average violation severity for LLaVA-72B on 4 GPUs.
Following the methodology reported in ModServe~\cite{modserve}, we define modality-specific SLOs as the P99 TTFT observed during isolated text-only, image-only, and video-only runs on vanilla vLLM.
The resulting TTFT SLOs are 0.53\,s (text), 0.69\,s (image), and 1.58\,s (video). We evaluate strict and relaxed performance objectives by scaling SLOs from 1$\times$ to 4$\times$.

We observe that \system consistently achieves the highest goodput by allowing more requests to meet their performance targets.
Goodput remains relatively stable across SLO scales, because it treats SLO attainment as a binary pass/fail outcome~\cite{revisiting-slo}, and it does not capture how severely requests miss their targets.
We therefore also report the percentage and severity of SLO violations.
As expected, both metrics decrease as the SLO scale increases, since requests are given more time to satisfy their latency objectives.
Across all SLO scales, \system consistently incurs fewer SLO violations than the disaggregation-based baselines and reduces violation severity by 2--3$\times$.
These results further confirm that \system meets its design objectives.

\takeaways{
Across the evaluation against the baselines, we observe that \system operates at the Pareto-optimal point, combining both high GPU utilization with high performance multimodal inference, compared to the rest of the baselines.
In addition, \system effectively reduces TTFT for text and keeps it low across both average and tail, by allowing them to bypass head-of-line blocking from heavier modalities.
At the same time, \system delivers sustained and balanced performance for images and videos, maintaining stable behavior without inflating tail latency. 

In contrast, disaggregation-based approaches suffer from resource fragmentation and very low GPU utilization.
They also exhibit pronounced tail amplification, while providing benefits primarily to multimodal models with heavy encoding overheads.
For models where encoding is not the dominant cost, disaggregation can instead hurt performance. 

These results highlight that improved scheduling is the key to achieving both responsiveness and efficiency in multimodal serving.
Augmenting disaggregated systems with \system consistently improves their performance, but does not have a significant impact on resource utilization, as it remains fundamentally constrained by the underlying resource partitioning of disaggregation.
}

\subsection{Ablation Study}
\label{ssec:ablation}

\begin{figure}[t]
    \centering
    \begin{subfigure}[t]{\linewidth}
        \centering
        \includegraphics[width=\linewidth]{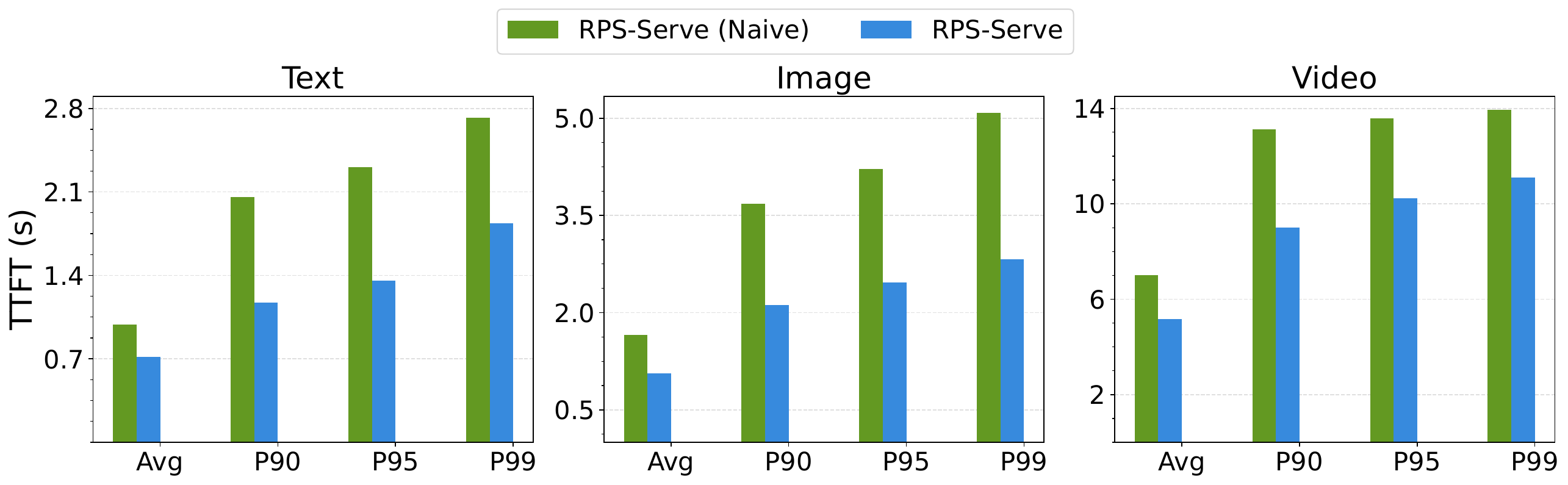}
        \caption{Request Classifier}
        \label{fig:ablation-classifier}
    \end{subfigure}
    \hfill
    \begin{subfigure}[t]{\linewidth}
        \centering
        \includegraphics[width=\linewidth]{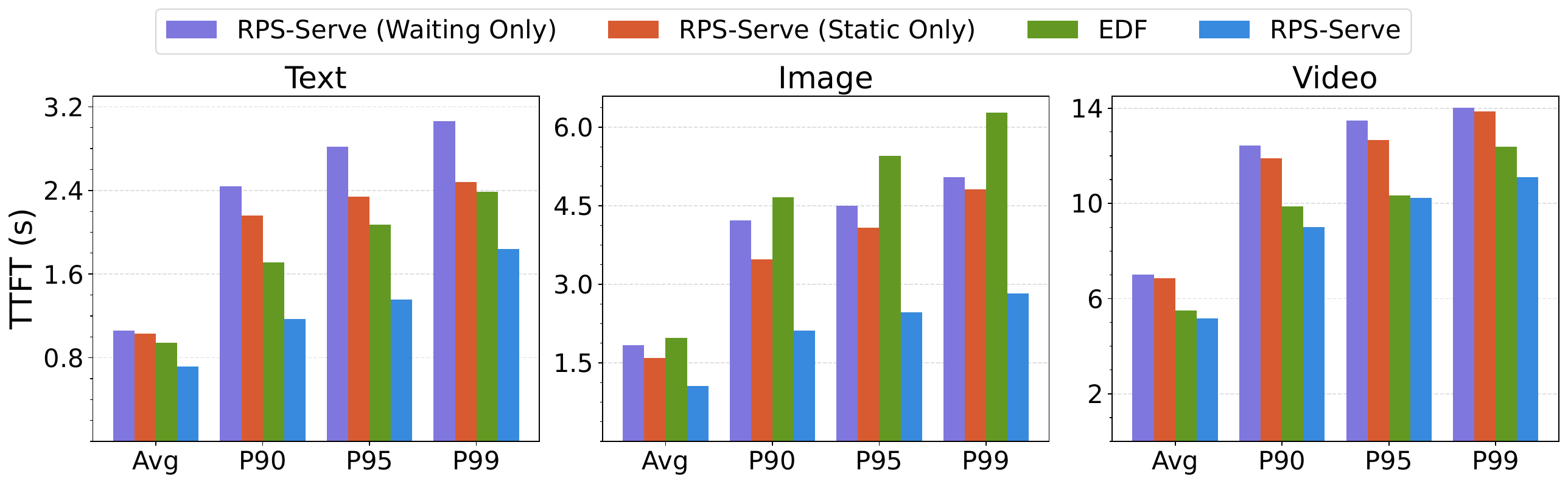}
        \caption{Priority Regulator}
        \label{fig:ablation-classifier-prio}
    \end{subfigure}
    \caption{Ablation study of the \system components.}
    \label{fig:ablation}
\end{figure}

Figure~\ref{fig:ablation} presents an ablation study comparing different design variants, reporting TTFT across modalities (text, image, video) under both average-case and tail behavior (P90, P95, P99) for LLaVA-7B on a single GPU. 

\customsection{Request Classifier} 
To highlight the importance of accurate request classification (Section~\ref{ssec:request-classifier}), we replace the Request Classifier of \system with a naive variant.
Naive classification directly maps text $\rightarrow$ sand, image $\rightarrow$ pebbles, and video $\rightarrow$ rocks, ignoring the actual GPU admission cost (e.g., latency and prefill time) of heterogeneous requests.
Figure~\ref{fig:ablation-classifier} shows that \system’s classifier significantly reduces TTFT across modalities, reducing latency by up to 1.8 seconds, on average, and up to 2.8 seconds at the P99 tail.
These improvements are particularly pronounced in the tail because naive misclassification may lead to severe scheduling inefficiencies. 
By incorporating model- and request-specific latency estimates, \system avoids over- or under-prioritizing requests that deviate from their modality class (e.g., very long text or short videos), as discussed in Figure~\ref{fig:clustering}.
Overall, these results demonstrate that accurate, cost-aware classification is critical for effective modality-aware scheduling, directly impacting average and tail performance. 

\customsection{Priority Regulator}
Priority-based scheduling (Section~\ref{ssec:priority-regulator}) assigns static priorities that favor lightweight sand over heavier pebbles and rocks, and augments them with an aging mechanism that accounts for waiting time and queue pressure to ensure progress for heavier requests. 
To isolate the contribution of these two priority terms, we compare \system against variants that use only static priorities, and only waiting-time-based prioritization, the latter resembling a shortest-wait-time or aging-only policy.

For completeness, we also compare against an Earliest Deadline First (EDF) baseline, a state-of-the-art priority-based scheduling approach in LLM serving systems that aims to minimize end-to-end latency~\cite{sola,tempo}. 
EDF assumes knowledge of each request’s deadline, or relies on prediction models to estimate output length and inference cost~\cite{s3,dynamollm,ranksched,intel-router}. 
This enables a direct comparison between \system and existing priority-driven schedulers, allowing us to evaluate the impact of modality-aware prioritization and aging in isolation.

Figure~\ref{fig:ablation-classifier-prio} shows that across all modalities, \system consistently achieves the lowest TTFT and significantly improves tail latency compared to all baselines, highlighting the importance of jointly incorporating both static and aging-based priority terms.
For text requests, the gains are most pronounced, particularly in the tail, where prioritizing lightweight requests allows them to complete earlier.
This results in reductions of approximately 1--1.5 seconds and yields an average TTFT of less than 1 second (700 ms), which is critical for meeting responsiveness targets in text-based applications~\cite{gpt4-sla, dist-serve, dynamollm}.
Furthermore, \system consistently outperforms EDF, as deadline-only prioritization fails to account for the significant variability in request characteristics.
As a result, EDF makes suboptimal scheduling decisions under heterogeneous workloads, whereas \system explicitly incorporates modality- and cost-aware prioritization to better handle such variability.

For image requests, EDF performs worse than \system, as it may prioritize requests with earlier deadlines but higher prefill cost, causing increased head-of-line blocking delays.
In contrast, \system's combined priority reduces TTFT by up to 3.5 seconds at the P99 tail compared to EDF.
For video requests, \system performs comparably to EDF but still achieves slightly lower TTFT (up to 1 second), indicating that even without explicit deadline information, its modality cost-aware and aging-driven prioritization allows heavier requests to complete within acceptable latency bounds.
Overall, \system’s priority formulation enables robust performance across heterogeneous multimodal workloads, outperforming even priority-based schedulers that rely on explicit deadline knowledge or prediction.

\subsection{Sensitivity Analysis}
\label{sec:sensitivity-analysis}

\begin{figure}[t]
    \centering
    \includegraphics[width=\linewidth]{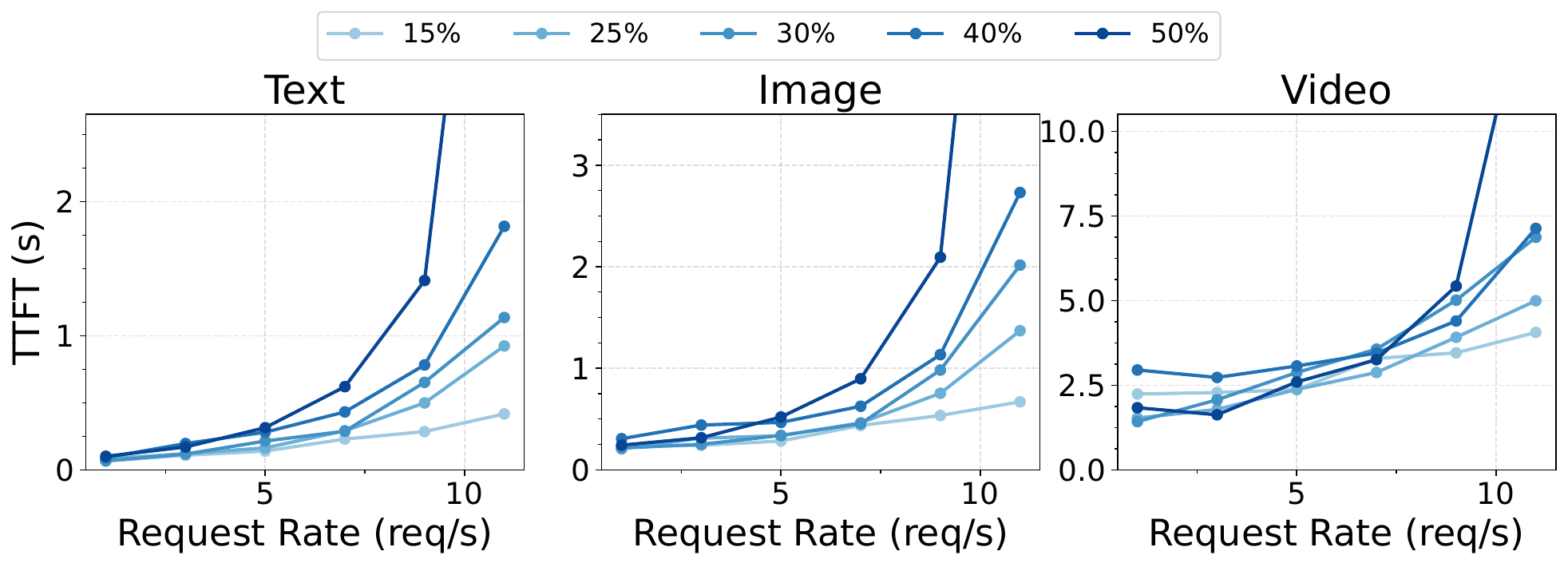}
    \caption{Performance sensitivity of \system to increasing request and multimodal load (\% of image + video).}
    \label{fig:sensitivity-workload}
\end{figure}

\begin{figure}
    \centering
    \begin{subfigure}[b]{0.49\linewidth}
        \centering
      \includegraphics[width=0.9\linewidth]{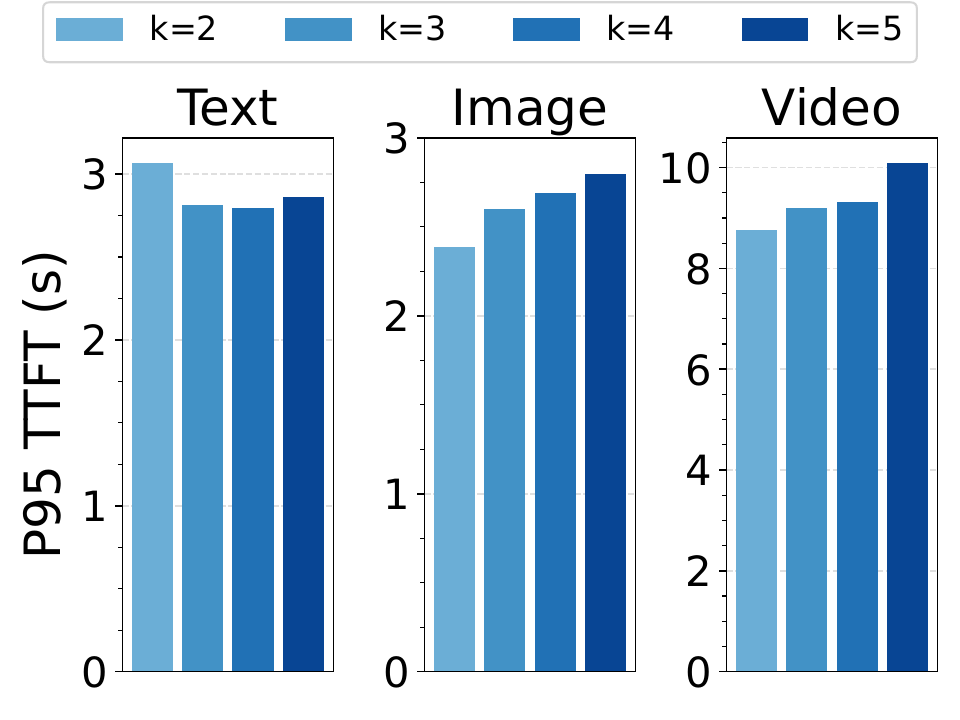}
      \caption{Performance Comparison}
      \label{fig:latency-gpus-qwen}
    \end{subfigure}
    \hfill
    \begin{subfigure}[b]{0.49\linewidth}
      \includegraphics[width=\linewidth]{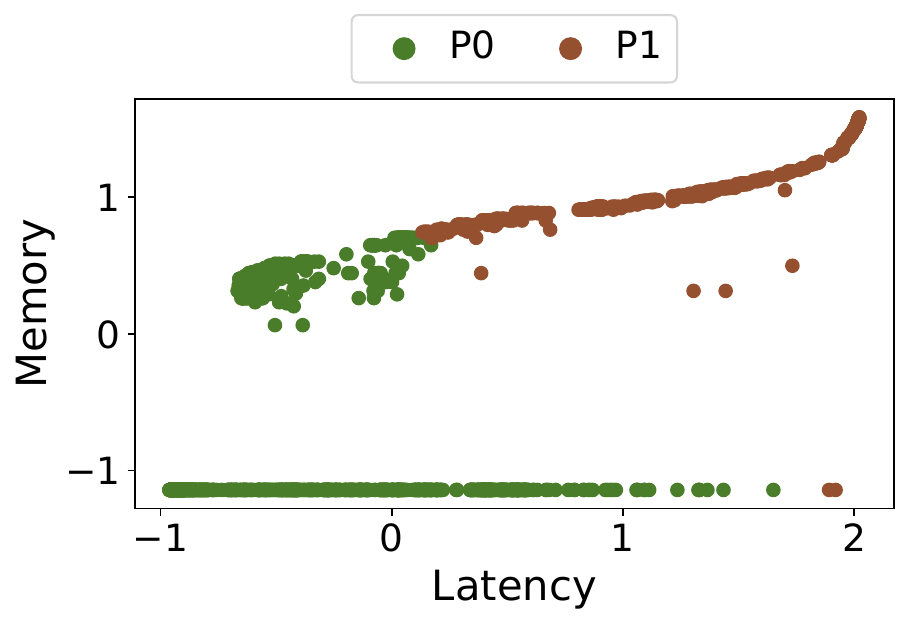}
      \caption{Clustering for k=2}
      \label{fig:k-2}
    \end{subfigure}
    
    \begin{subfigure}[b]{0.49\linewidth}
      \includegraphics[width=\linewidth]{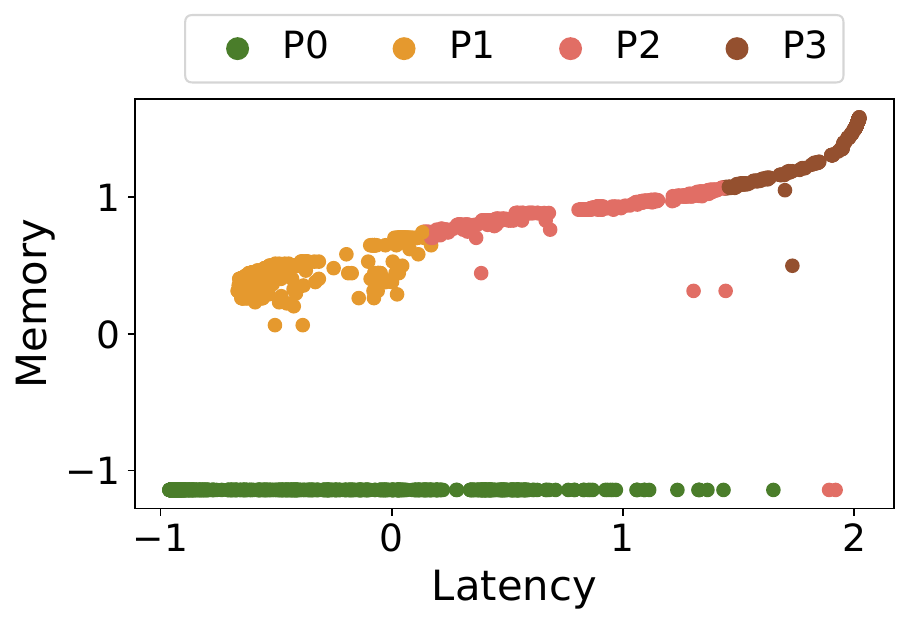}
      \caption{Clustering for k=4}
      \label{fig:k-4}
    \end{subfigure}
    \hfill
    \begin{subfigure}[b]{0.49\linewidth}
      \includegraphics[width=\linewidth]{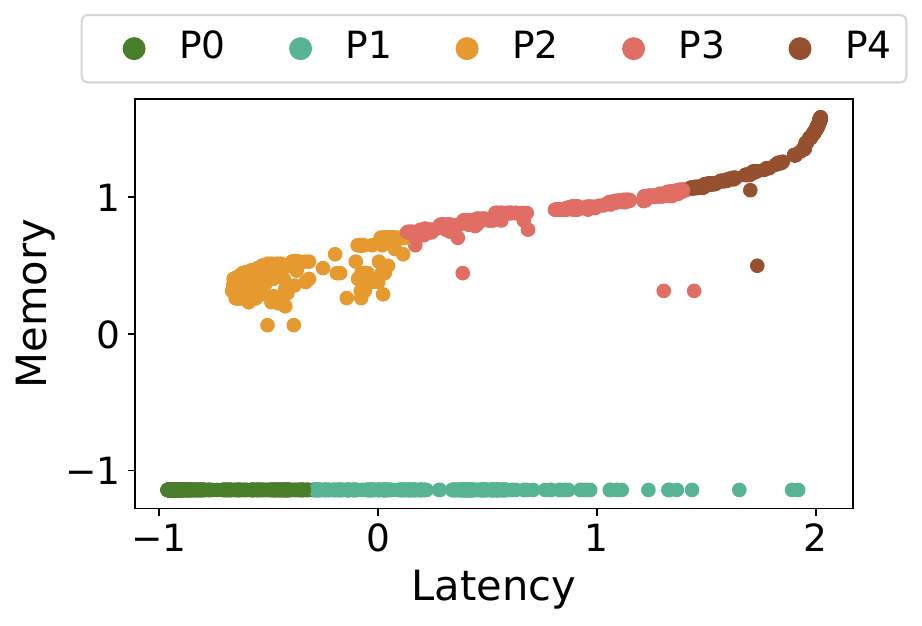}
      \caption{Clustering for k=5}
      \label{fig:k-5}
    \end{subfigure}
    
    \caption{Performance sensitivity of \system to increasing number $k$ of request classes.}
    \label{fig:sensitivity-clustering}
\end{figure}

\customsection{Varying Workload}
Figure~\ref{fig:sensitivity-workload} shows how TTFT evolves across text, image, and video requests as both the request rate and the fraction of image and video requests increase for LLaVA-7B on a single GPU.
As the workload becomes more multimodal-heavy, TTFT rises across all modalities, particularly at high load.
Nevertheless, text requests consistently achieve the lowest TTFT, while image and video requests experience higher latency, reflecting the differentiated scheduling enabled by the Rocks-Pebbles-Sand abstraction.
For reference, the remainder of the evaluation used the workload with 30\% of image and video requests combined (Section~\ref{ssec:experimental-setup}).

\customsection{Number of Request Classes}
By design, \system partitions requests into three classes, reflecting the Rocks-Pebbles-Sand abstraction and the insights from the multimodal workload characterization (Section~\ref{ssec:characterization-iso}).
We evaluate this design choice by varying the number of request classes \(k\) from 2 to 5.
This comes with changes in the system configuration, since it means adjusting the number of clusters in the Request Classifier, the number of queues in the Queue Manager, and the per-class parameters in the Priority Regulator, accordingly.
Figure~\ref{fig:latency-gpus-qwen} reports P95 TTFT for Qwen-27B on a single GPU.

With \(k=2\), text requests experience the highest latency, indicating insufficient differentiation between lightweight and resource-intensive requests.
Increasing to \(k=3\) substantially reduces text TTFT with only a modest increase for image and video requests.
Further increasing \(k\) beyond 3 yields no additional benefit for text while increasing latency for image and video requests.
Figures~\ref{fig:k-2}--\ref{fig:k-5} explain this behavior by showing the corresponding cluster assignments, complementing Figure~\ref{fig:clustering}.
With \(k=2\), text and image requests share the same high-priority class.
With \(k=4\), video requests are further split into short and long videos, providing no benefit to the higher priority requests, while \(k=5\) additionally separates text requests by prompt length, further delaying image and video requests.
Overall, \(k=3\) provides the best balance between request differentiation, performance and system configuration complexity, further validating the design of the Rocks-Pebbles-Sand abstraction.

\subsection{Discussion and Future Work}
\label{sec:discussion}

While \system significantly improves multimodal inference performance, it currently supports only text, image, and video modalities.
Our sand--pebbles--rocks abstraction is general enough to extend to other modalities (e.g., audio or 3D data), although doing so may require retraining the classifier and recalibrating the priority regulator.
The design of the Queue Manager also provides a natural hook for future extensions.
For example, it could support per-class admission control or SLO-aware reordering within each queue without requiring changes to the classification or estimation components.
More broadly, future work includes supporting multimodal output generation (e.g., image or video responses) and extending \system beyond MLLMs to any-to-any multimodal models.
\section{Related Work}
\label{sec:related-work}

\custompar{Multimodal LLM inference}
State-of-the-art multimodal serving systems use disaggregation to reduce the head-of-line blocking delays introduced by vision encoding.
ModServe~\cite{modserve} separates the vision encoder from the LLM backend, isolating expensive encoding operations from text processing.
EPD~\cite{epd} further decomposes execution into independent encoding, prefill, and decode stages, enabling stage-specific optimizations.
ElasticMM~\cite{elasticmm} generalizes this approach through hierarchical disaggregation, separating workloads both by modality and by execution stage while dynamically adjusting resource allocation and parallelism.
Collectively, these systems reduce cross-stage interference by dedicating resources to the individual stages of the multimodal inference pipeline. 
Orthogonal to those, other works have tried to accelerate multimodal LLM inference by revisiting the attention mechanism of MLLMs to reduce computation~\cite{boostingmllm} or by caching only relevant tokens~\cite{infmllm,elasticcache}.
These solutions are model-specific and they can potentially reduce model accuracy. Lastly, ServeGen~\cite{servegen} just characterizes large-scale LLM serving workloads and introduces a workload generator, but focuses mainly on language and image-text models, treating video as image bursts.  

\custompar{LLM inference} 
Several works address head-of-line blocking, that is typically caused by long prefills in traditional LLM workloads.
Chunked-prefill~\cite{sarathi,sarathi-serve,vllm-chunk} is now a standard approach offered by default in vLLM~\cite{vllm-chunk}.
Pipelined execution~\cite{orca,loongserve,alpaserve} aims to overlap prefill and decode phases, while KV-cache compression~\cite{cachegen,scissorhands} and offloading strategies~\cite{flexgen,attentionstore} reduce memory pressure. 
These methods improve responsiveness for long-text scenarios but do not generalize to multimodal inference, where preprocessing and encoding of visual inputs dominate time-to-first-token latency.

Other works have tried to improve LLM inference by predicting the output length or the workload impact of requests using predictor and cost models~\cite{intel-router,s3,sola,seqlenpred,ranksched,responselen,dynamollm}.
Other approaches have tried to use an SLO-based approach to accelerate interactive requests and reduce latency~\cite{dist-serve,mooncake,andes,sola,tempo,fastserve,sloawaresched}.
Dynamic batching and fairness-oriented policies have been proposed for LLM serving~\cite{llumnix,dynamollm,fairsched}.
All of the above techniques and methods target homogeneous text workloads and ignore the extreme heterogeneity of multimodal requests.
Our approach complements these works by introducing modality-aware prioritization.
\section{Summary}
\label{sec:conclusion}

This paper introduces \system, a modality-aware scheduling framework for resource-efficient, high-performance multimodal inference.
\system leverages a simple yet effective abstraction (\emph{rocks, pebbles, and sand}) to capture the varying resource demands of video, image, and text requests, and combines cost-aware classification with adaptive priority scheduling to balance responsiveness and utilization.
Evaluation demonstrates faster inference and improved GPU utilization compared to disaggregation-based multimodal serving systems.
\system is open source to enable reproducible research and community contributions.

\begin{acks}
We thank the anonymous reviewers for their constructive feedback and our shepherd for their guidance.
The work was partially funded by the Madrid Regional Government through the César Nombela grant (2024-T1/COM-31302) and by the Comunidad de Madrid through the DATIA project, co-funded by the European Union's FEDER funds.
The work was also supported by grant PID2022-142290OB-I00, funded by MCIN/AEI/10.13039/501100011033 and FEDER, UE, and by grant CEX2024-001471-M, funded by MICIU/AEI/10.13039/ 501100011033. 
\end{acks}

\bibliographystyle{ACM-Reference-Format}
\bibliography{paper}
\balance

\end{document}